# Optimization-Based Thickness Estimation and Depth-Resolved FIB-Induced Damage Characterization via Multislice Electron Ptychography

Keun-Yeol Park[1], Chuqiao Shi[2], Murat Tuna Pamuk[2], Sooyoung Cheong[1], Seonyu Lee[1], Juhui Oh[1], Yu-Tsun Shao[2,3], and Celesta S. Chang[1,4,*]

[1] *Department of Physics and Astronomy, Seoul National University, Seoul 08826, Korea*

[2] *Mork Family Department of Chemical Engineering and Materials Science, University of Southern California, Los Angeles, CA 90089, USA*

[3] *Core Center of Excellence in Nano Imaging, University of Southern California, Los Angeles, CA 90089, USA*

[4] *Institute of Applied Physics, Seoul National University, Seoul 08826, Korea*

* *Corresponding author: celesta@snu.ac.kr*

## ABSTRACT

Quantitative electron microscopy requires accurate knowledge of specimen thickness because dynamical scattering strongly affects image contrast and diffraction intensities. In multislice electron ptychography (MEP), specimen thickness is a required input parameter to its forward model, yet the local thickness of a transmission electron microscope (TEM) lamella is often precisely what is unknown. Moreover, thickness is coupled to probe defocus and the number of slices, making its determination from the reconstruction itself nontrivial. Here, we assess whether joint Bayesian optimization of these parameters can provide a physically meaningful estimate of local thickness from four-dimensional scanning transmission electron microscopy (4D-STEM) data. Applied to a wedge-shaped silicon lamella, the optimization reproduced similar local thickness variation measured by electron energy loss spectroscopy (EELS), with a systematic offset of 10–17% relative to the EELS estimates. Using the optimized reconstruction parameters, depth-resolved MEP further separated the crystalline silicon interior from focused ion beam (FIB)-induced amorphous surface layers within the same reconstructed volume. Regions milled at final voltages of 2, 5, 8, and 30 kV yielded amorphous-layer thicknesses of 3.2, 5.0, 7.0, and 26.4 nm, respectively, increasing monotonically with milling voltage and agreeing reasonably with previous cross-sectional measurements. These results show that applying MEP coupled with Bayesian optimization on a single 4D-STEM dataset can provide both local thickness estimates and depth-resolved characterization of FIB-induced damage, offering a route toward more self-consistent, thickness-aware quantitative 4D-STEM analysis.

## 1. Introduction

Specimen thickness is a governing experimental variable in transmission electron microscopy (TEM) and scanning transmission electron microscopy (STEM), as it determines the extent of electron scattering before an image or diffraction pattern is formed. In crystalline specimens, dynamical scattering causes the intensity and phase of the exit wave to depend strongly and nonlinearly on the propagation distance [1,2]. Consequently, specimen thickness directly influences both the image contrast and the diffracted intensities [3]. This dependence is especially important for atomic-resolution imaging and quantitative analysis, including aberration-corrected STEM measurements of individual atomic columns and four-dimensional STEM (4D-STEM) analyses based on subtle intensity variations in convergent-beam electron diffraction (CBED) discs. In 4D-STEM, these variations can be used to identify structural phases, determine local symmetry, quantify polar distortions, and map ferroelectric domains [4–6]. Changes in specimen thickness can generate scattering signals comparable to those arising from the structural features of interest, potentially leading to inaccurate or even ambiguous physical interpretations when the local thickness is not known [7–9].

This requirement presents a practical challenge for site-specific TEM specimens prepared by focused ion beam (FIB) milling. The total thickness of a FIB-prepared lamella is determined by the milling process and generally varies spatially, particularly in wedge-shaped specimens [10]. Moreover, the Ga ion beam implants ions, damages the crystalline structure, and amorphizes material near the milled surfaces [11]. Although progressively lower milling voltages are routinely used to reduce such damage [12–16], a FIB-prepared lamella generally consists of a crystalline bulk bounded by amorphous surface layers. These layers overlap the crystalline bulk along the electron-beam direction and may not be readily distinguished in conventional TEM images, despite their contribution to electron scattering and the resulting dynamical and nonlinear imaging effects [17]. For quantitative analyses sensitive to dynamical scattering, the total local thickness alone may therefore be insufficient: the fraction of the lamella occupied by crystalline and amorphous material, and their distribution along the beam direction, can also influence the measured signal. FIB-induced damage has traditionally been characterized by cross-sectional imaging of specimens prepared specifically for damage-depth measurements [18]. Such measurements, however, require a separately prepared specimen which is cut across the lamella. Thus, neither the local total thickness nor its depth-dependent structural crystallinity has been readily accessible from the lamella of interest.

Multislice electron ptychography (MEP) provides a framework for examining both quantities from the same 4D-STEM dataset. By explicitly treating dynamical scattering within the forward model, MEP reconstructs the specimen as a stack of slices along the beam direction, with sub-Ångström lateral resolution [19] and nanometer-scale depth resolution [20,21]. Structures that overlap in projection, such as amorphous layers and the crystalline interior, can therefore be separated within a single reconstructed volume. MEP reconstruction, however, requires the total specimen thickness to be specified as a model parameter to begin with, even though its local value is not generally known in advance. Moreover, thickness cannot be determined independently, because the reconstruction outcome also depends on coupled parameters such as probe defocus and the number of slices. Reconstruction fidelity depends on these intertwined parameters [22], and Bayesian optimization (BO) has been introduced as a systematic framework for determining them [23,24]. This raises an important question: does the thickness selected by optimization merely represent a reconstruction parameter that produces the best numerical result, or can it be interpreted as a measure of the physical specimen thickness?

Here, we address this question using a wedge-shaped silicon TEM lamella containing four distinct regions milled at final accelerating voltages of 2, 5, 8, and 30 kV (**Fig. 1a**). From each region, we acquired a 4D-STEM dataset and reconstructed it by using MEP (**Fig. 1b**). We first used BO to jointly optimize the total specimen thickness, the probe defocus, and the number of slices, and examined whether the optimized thickness can be interpreted as an estimate of the physical lamella thickness. Utilizing the optimized parameters, we then performed a depth-resolved reconstruction to separate the FIB-induced amorphous layers from the crystalline bulk and quantified their thicknesses as a function of milling voltage. This approach provides a model-based estimate of local specimen thickness together with depth-resolved structural information from the same 4D-STEM dataset, without requiring a separately prepared cross-sectional specimen. Beyond FIB damage characterization, this capability establishes a route toward thickness-aware quantitative electron microscopy where locally estimated specimen thickness can constrain simulations and the interpretation of measurements strongly affected by dynamical scattering.

## 2. Methods

### 2.1. TEM specimen preparation with FIB.

A cross-sectional silicon TEM lamella was prepared from a commercial Si (001) wafer using a Thermo Fisher Scientific Helios G4 dual-beam FIB-SEM system with a Ga ion source, with the lamella normal parallel to the [110] zone axis. Prior to FIB loading, carbon-based protection layers were applied to the sample surface using a permanent yellow marker [25]. Following the conventional FIB lift-out and milling protocols [10,11], the lamella was initially thinned down to approximately 200 nm at an accelerating voltage of 30 kV. To systematically vary the extent of FIB-induced damage, the lamella was divided into four regions with a lateral width of approximately 4 μm, with final milling voltages of 2, 5, 8, and 30 kV. To minimize the curtaining effect, all regions underwent a rectangular polishing step at 30 kV with a ±3° over-tilt, reaching a thickness of approximately 100 nm for the regions subsequently polished at low voltages and approximately 60 nm for the region finalized at 30 kV. The 8 and 5 kV regions were then polished at the corresponding voltage with the same ±3° over-tilt to a target thickness of approximately 30 nm. For the region finalized at 2 kV, an intermediate 5 kV polishing step first reduced the thickness to approximately 60 nm, followed by a final 2 kV polish to approximately 30 nm. The detailed milling parameters, including beam currents, targeted thicknesses, and scan modes, are summarized in **Table S1**.

### 2.2. 4D-STEM data acquisition and multislice ptychographic reconstruction.

4D-STEM datasets for multislice electron ptychography were acquired using a Thermo Fisher Scientific Spectra 200 S/TEM operated at 200 kV equipped with a cold field emission gun (X-CFEG), a probe spherical aberration ($C_s$) corrector, and a Dectris ARINA direct electron detector. Data were acquired with a probe convergence semi-angle of 30 mrad, a real-space scan step size of 0.36 Å, and a reciprocal-space sampling of 0.81 mrad/pixel, yielding a maximum collection angle of approximately 78 mrad. The dwell time was 35 μs per probe position, resulting in a total incident electron dose of approximately $7 \times 10^4$ $e^-/Å^2$. The probe defocus was set by positioning the crossover approximately 30 nm above the center of the specimen, resulting in overfocus for the 2, 5, and 8 kV regions whereas the crossover lay within the thicker 30 kV region corresponding to underfocus. Each dataset comprised 256 × 256 scan pixels, corresponding to a 9.2 × 9.2 nm² field of view, and was pre-processed using standard dark-reference subtraction and gain normalization.

Multislice electron ptychographic reconstructions were performed using the least-squares maximum-likelihood multislice algorithm [20,26,27] implemented in the GPU_MS engine of the fold_slice package [28], which builds upon PtychoShelves [29]. The implementation incorporating a tilt propagator [21,30] was used to account for specimen misorientation in the forward model [31]. Specimen tilt angles were determined from the position and shift of Kikuchi bands relative to the [110] zone axis in the CBED patterns. The calculated diffraction intensity was convolved with a Gaussian with a standard deviation of one detector pixel.

For parameter estimation, a region of interest (ROI) of 64 × 64 scan pixels, corresponding to 2.3 × 2.3 nm², was cropped from each dataset. The total specimen thickness, the probe defocus, and the number of slices were jointly optimized within this ROI by BO as described in Section 2.3. The optimized parameters were then used to reconstruct the full 256 × 256 scan pixel dataset for 1000 iterations. Since the optimized thickness represents the local thickness averaged over the optimization ROI and the wedge-shaped lamella varies in thickness across the field of view, the depth-resolved analysis and three-dimensional (3D) visualizations were restricted to a $4.6 \times 4.6$ nm² region centered on the optimization ROI. The selected regions (**Fig. S1**) and the near-surface layer structures in the lamella (**Fig. S2**) are shown in the Supplementary Information, with the experimental and reconstruction parameters listed in **Table S2**.

### 2.3. Bayesian optimization for parameter estimation.

Bayesian optimization (BO) was used to jointly determine three parameters required for multislice electron ptychography: the total specimen thickness $t$, the probe defocus $\Delta f$, and the number of slices $N$, with the slice spacing defined as $\Delta z = t/N$. The optimization workflow was adapted from the approach of Cao et al. [23] and reimplemented in Python using BoTorch 0.16.1 [32]. Before performing BO, the scan step size and scan rotation were calibrated for each dataset by separate one-parameter sweeps. These parameters, together with the specimen tilt angles determined from the CBED patterns, were held fixed during the optimization.

Search bounds for $t$, $\Delta f$, and $N$ were set separately for each dataset to bracket the expected values, and each candidate parameter set was evaluated by a reconstruction with a fixed number of 100 iterations. This differs from the wall-clock time limit used by Cao et al. [23] which is useful when the candidate parameters differ widely in computational cost. The reconstruction error $\varepsilon$ was defined as the root-mean-square difference between measured and simulated diffraction amplitudes, and the BO objective was defined as $Y = -\log \varepsilon$ such that maximizing $Y$ minimizes $\varepsilon$. To prevent other model degrees of freedom from compensating for an incorrect thickness, multiple probe states and orthogonal probe relaxation were disabled during optimization, and all remaining reconstruction settings were held fixed. A Gaussian-process (GP) surrogate was fitted to the observed values of $Y$. Because $N$ is integer-valued, it was handled by a rounding transform applied within the covariance function, so that the surrogate is constant between adjacent integer values [33]. Candidate parameter sets were selected using the q-batch upper confidence bound (qUCB) acquisition function [34], a parallel generalization of $\alpha(x) = \mu(x) + \sqrt{\beta}\cdot\sigma(x)$ [35], with $\mu$ and $\sigma$ being the posterior mean and standard deviation of the surrogate respectively, and $\beta$ the exploration weight. The optimization settings and the dataset-specific search bounds are listed in **Table S3**.

### 2.4. Depth-resolved characterization and quantification of the amorphous-layer thickness.

Three-dimensional (3D) renderings of the reconstructed volumes were generated in Tomviz by mapping the reconstructed phase to color and opacity. Regions with higher phase values, corresponding to crystalline silicon, were rendered with greater opacity. FIB-induced amorphous-layer thickness was quantified from phase-versus-depth profiles of individual atomic columns. Atomic column positions were identified from the central slice of each reconstructed volume using the peak-finding algorithm in *kemstem* [36]. At each column position the reconstructed phase was averaged over a 3 × 3 pixel region of the reconstructed object in every slice to generate a one-dimensional phase profile along the beam direction. Each profile was fitted with a symmetric double error function, $A[\mathrm{erf}(B(x - x_1)) + \mathrm{erf}(B(x_2 - x))]$, where $A$ is the amplitude, $B$ is the transition steepness, and $x_1$ and $x_2$ denote the positions of the two amorphous–crystalline interfaces, corresponding to the points where each error function reaches half of its own step height. The crystalline thickness $t_c$ is the distance between the two fitted interfaces $t_c = x_2 - x_1$. For a reconstructed volume of total optimized thickness $t*$, the combined amorphous thickness was therefore $t^* - t_c$ and the mean thickness of a single amorphous layer $t_a$ was calculated by $(t^* - t_c)/2$. This analysis was applied to all atomic columns within the 2.3 × 2.3 nm² optimization ROI. Profiles for which the fit failed to resolve two distinct interfaces were excluded by the criterion $B(x_2 - x_1) < 2$. The mean of $t_a$ over the remaining columns was taken as the amorphous-layer thickness, with error bars spanning the minimum and maximum of the individual column estimates.

### 2.5. Independent thickness measurement by electron energy-loss spectroscopy.

Independent measurements of the local specimen thickness were performed by electron energy-loss spectroscopy (EELS) using a $C_s$-corrected monochromated scanning transmission electron microscope (Themis Z, Thermo Fisher Scientific) equipped with a Gatan imaging filter (Quantum ER965). EELS data were acquired at an accelerating voltage of 300 kV with a 2.5 mm aperture, a camera length of 29.5 mm, and an energy dispersion of 0.5 eV/channel. The spectra were collected in a single scan pass with a step size of 0.5 nm and a pixel dwell time of 5 ms. The acquired data were analyzed using Gatan Microscopy Suite (GMS) software. The specimen thickness was estimated from the EELS spectra using the log-ratio method, where the relative thickness ($t/\lambda$) was determined by the ratio of the total integrated intensity ($I_t$) to the zero-loss peak intensity ($I_0$). To determine the absolute thickness, the inelastic mean free path ($\lambda$) for Si was calculated as 138 nm using the parameterization of Malis et al. [37]. A single value of $\lambda$ was used for the amorphous and crystalline regions of the lamella, since the two have been reported to differ by less than 1% under comparable conditions [38].

## 3. Results

### 3.1. Optimization-based estimation of local specimen thickness.

The first requirement for quantitative analysis of the reconstructed lamella is to determine its local thickness. In multislice ptychographic calculations, specimen thickness is treated as an input to the forward model. Here, however, the thickness of the specimen is precisely the quantity that is unknown. This creates an apparent circular problem in that we seek to determine the thickness from the very reconstruction that requires it. Furthermore, thickness cannot be varied independently in this search. Changing the thickness alters the propagation of the electron wave through the specimen, while the probe defocus changes the incident wavefield and the number of slices determines how that propagation is discretized. Therefore, adjustments to probe defocus or slice number can partially compensate for the effect of an incorrect thickness on the calculated diffraction pattern. Consequently, a low reconstruction error obtained at a particular assumed thickness does not by itself establish that the thickness is physically correct.

We therefore determine the total thickness $t$, the probe defocus $\Delta f$, and the number of slices $N$ jointly and examine whether the resulting optimum is uniquely constrained by the experimental diffraction data. To perform this search, we coupled multislice ptychographic reconstruction with a Bayesian optimization engine described in Section 2.3 (**Fig. 2a**). The central question here is not simply which parameter set $X^* = [t^*, \Delta f^*, N^*]$ gives the lowest reconstruction error, but whether the resulting optimum $t^*$ can be interpreted as the physical thickness of the specimen rather than merely as a model setting that best reproduces the diffraction data. In our work, each candidate was reconstructed for a fixed number of 100 iterations as the reconstruction error largely plateaued within this range (**Fig. 2b**), while the ranking among parameter sets remained consistent beyond this point (**Fig. S3**). Full convergence of every candidate was therefore unnecessary as their relative ranking was preserved. The 2 kV region was used below as a representative case, after which the same procedure was applied to the 5, 8, and 30 kV regions.

Joint optimization of the three parameters on the 2 kV dataset converged to an optimized thickness of $t^*$ = 245 Å, with the corresponding probe defocus $\Delta f^*$ = −207 Å and the number of slices $N^*$ = 20. Plotting the error of every evaluated candidate against each parameter individually (**Fig. 2c**) gives a single well-defined minimum within the search bounds for all three parameters. The scatter above the lower envelope in each projection arises because the other two parameters vary simultaneously. Importantly, no competing minimum of comparable error was observed within the explored parameter space, indicating that the retrieved thickness belongs to a uniquely favored joint solution rather than to one of multiple equivalent parameter combinations.

Mapping the error over the thickness–defocus plane at fixed $N^* = 20$ shows that the error-minimizing thickness follows a sloped line rather than a horizontal one (**Fig. 2d**). This indicates that fixing the defocus at an incorrect value would shift the retrieved thickness. This coupling is physically expected because both parameters shift the axial position of the reconstructed wavefield. The thickness $t$ sets the distance over which the wave is propagated, while defocus $\Delta f$ enters the forward model as a phase quadratic in the diffraction-plane coordinate [1]. Variations in these two parameters can therefore partially compensate for one another during reconstruction. The number of slices $N$ is also coupled to both parameters because it determines the axial discretization $\Delta z = t/N$. When the optimization was repeated with N held fixed at each value (**Fig. 2e**), the error reached a minimum near $N^* = 20$, whereas the corresponding optimized thickness and defocus increased monotonically with $N$. Thus, fixing any one of the three in advance can bias the other two, providing strong grounds for their joint optimization. For $t^*$ to serve as a meaningful estimate of specimen thickness, we next examined how the reconstruction depends on $N$, and whether changes in this reconstruction-dependent parameter alter the optimized thickness $t^*$.

To assess the robustness of the optimized thickness and slice number to axial sampling and layer regularization, we performed two complementary tests. First, with $t^*$ and $\Delta f^*$ held fixed, we varied only $N$ and extracted phase-versus-depth profiles along a single silicon column to measure the 10–90% transition width at the two amorphous–crystalline interfaces (**Fig. S4**). The transition width falls steeply up to $N = 18$, changes by only a few Ångströms at $N = 20$ and 22, and remains nearly constant thereafter. Similar stability of MEP depth sectioning with decrease in slice spacing, once sufficient axial sampling is reached, has been reported previously [21]. In our data, $N^* = 20$ lies near the point beyond which finer slicing no longer improves depth resolution. Second, we repeated the full optimization to confirm the stability of the optimized $t^*$ while varying the layer-regularization strength—a dimensionless setting that suppresses object components varying rapidly along the beam direction—from 0.1 to 0.5 (fixed at 0.2 elsewhere in this work). Across this range, the optimal slice number $N^*$ shifts from 20 to 24, with $\Delta z$ correspondingly falling from 12.3 to 10.3 Å, whereas the optimized total thickness remains within 3% of $t^* = 245$ Å (**Fig. S5**). Together, these tests show that $N^*$ depends on the axial sampling and regularization of the reconstruction, whereas the retrieved thickness $t^*$ remains substantially more stable to these settings.

Robustness to reconstruction settings alone, however, does not establish that $t^*$ corresponds to the physical specimen thickness. If $t^*$ represents a physical property of the specimen, it should reproduce real spatial variations in thickness independently of the reconstruction procedure. We tested this using the wedge geometry of the 2 kV region. Because the optimization can be applied independently to cropped portions of the same dataset, the top, middle, and bottom subregions were optimized separately (**Fig. 3a, b**). The resulting thicknesses were $t^*$ = 23.2, 24.5, and 26.5 nm, respectively, increasing from top to bottom as expected for the wedge-shaped geometry formed by ion milling (**Fig. 3c**). The local thickness along the same direction was then measured independently by the EELS log-ratio method (**Fig. 3d, e**), yielding 27.9, 28.7, and 29.4 nm for the depths corresponding to the top, middle, and bottom subregions. Both methods reproduce the same spatial trend, although the optimized thicknesses are smaller than the EELS values by 17, 15, and 10%, respectively. The absolute EELS-derived thickness scales directly with the assumed mean free path $\lambda$, for which the parameterization used here has a quoted accuracy of approximately ±20% [37]. The systematic offset, the physical interpretation, and limitations of the optimized thickness are considered further in Section 4. Nevertheless, its ability to reproduce the local thickness variation within a single lamella, together with its robustness against the reconstruction settings tested above, verifies the interpretation of $t^*$ as an estimate of the local specimen thickness rather than solely as a numerical reconstruction parameter.

Finally, applying the same joint optimization to the 5, 8, and 30 kV regions produced a single, well-defined optimum within the explored parameter space for each dataset (**Fig. S6**). The optimized thickness, defocus, and number of slices for all four regions are summarized in **Table S2**. These parameters were then used to reconstruct the full datasets, providing the depth-resolved volumes analyzed in Section 3.2 to quantify the FIB-induced amorphous layers.

### 3.2. Depth-resolved characterization of FIB-induced damage.

Having established the reconstruction parameters and local specimen thicknesses in Section 3.1, we next use the depth-resolved MEP reconstructions to determine how the FIB-induced damage is distributed through the lamella as a function of the final milling voltage. The beam current was not varied as an independent parameter, since the accelerating voltage is the primary parameter governing the ion penetration depth and the resulting thickness of the FIB-induced amorphous layer [39]. Because the amorphous surface layers and crystalline silicon bulk overlap along the electron-beam direction in conventional projection images, separating them requires resolving their structure along the depth direction. More specifically, locating the amorphous–crystalline interfaces requires tracking the persistence of periodically ordered Si atomic columns across successive depth slices, rather than relying only on an overall change in image contrast. This analysis therefore requires atomic lateral resolution throughout the reconstructed volume. The reconstruction of all four regions resolved the atomic lattice at sub-Ångström resolution (**Fig. S7**), allowing crystalline order to be tracked through the reconstructed volume.

The reconstruction directly reveals the transition from the FIB-induced amorphous layers to the crystalline bulk. The rendered volume of the 5 kV region is shown as a representative example in **Fig. 4a**, where vertical crystalline columns are bounded by disordered surface regions along the beam direction on both sides. Lateral slices extracted at successive depths trace this transition in the sample plane. The reconstruction showed disordered contrast of amorphous silicon near the surface (**Fig. 4b**), atomic columns emerging from the disordered background at the amorphous–crystalline interface (**Fig. 4c**), and a fully ordered lattice at the bulk crystalline (**Fig. 4d**). The same progression was observed in the other three regions, milled at 2, 8, and 30 kV (**Fig. S8**). Notably, the region milled at the lowest voltage of 2 kV exhibits a distinctively disordered surface region in its outermost slices, showing the capability of ptychographic reconstruction to resolve the thin amorphous layer that low-voltage milling leaves behind.

To quantify these transitions, we extracted phase-versus-depth profiles at individual atomic columns. Since the reconstructed phase reflects the projected electrostatic potential within each slice [20], an aligned atomic column produces a strong phase signal within the crystalline region that diminishes as crystalline order is lost toward the specimen surfaces. This behavior is illustrated for a silicon dumbbell in the 5 kV region (**Fig. 4e, left panel**), where the two silicon columns appear as continuous high-phase features through the crystalline bulk and weaken toward both surfaces. The right panel of **Fig. 4e** gives the corresponding depth profile at one of these columns. Within a crystalline atomic column, the phase contribution from each Si atom at successive depths can be approximated by a Gaussian peak. Because these atoms are closely spaced along the beam direction, their summed phase profile forms a high, nearly uniform plateau within the crystalline core. Toward amorphous surfaces, the loss of periodic atomic column alignment reduces the signal to a low background level. The resulting profile was fitted with a symmetric double error function as described in Section 2.4. The fitted interface positions, $x_1$ and $x_2$, define the crystalline thickness $t_c = x_2 - x_1$, as illustrated in **Fig. 4e**. Combined with the optimized total thickness $t^*$, this value yields the thickness of a single amorphous layer, $t_a = (t^* - t_c)/2$. Applying this analysis to individual atomic columns across each region provided spatially resolved estimates of $t_c$ and $t_a$ (**Fig. S9** and **S10**).

The resulting amorphous-layer thicknesses $t_a$ are summarized in **Fig. 5**. The thickness of a single amorphous layer increases with the final milling voltage, from 3.2 nm at 2 kV to 5.0 nm at 5 kV, 7.0 nm at 8 kV, and 26.4 nm at 30 kV. This trend is consistent with the reduced extent of amorphization expected from sequential reduction of the milling voltage. Error bars indicate the full range of values obtained from the individual atomic column profiles within the analyzed region. The symmetry of the two layers was assessed by comparing the top layer, $x_1$, with the bottom layer, $t^* - x_2$ in each region. Their absolute difference, $|x_1 + x_2 - t^*|$, was 0.7, 0.3, 0.3, and 0.7 nm at 2, 5, 8, and 30 kV, respectively, with all values less than one slice spacing. Profiles for which two interfaces could not be reliably separated were excluded according to the criterion described in Section 2.4, removing fewer than 2% of the columns in the 2, 5, and 8 kV regions and 16% for the 30 kV region.

To place these estimates in context, **Fig. 5** compares the present results with amorphous-layer thicknesses previously determined from separately prepared cross-sectional specimens [11–13,16,18]. Care is required in making this comparison because different measurements use different definitions of the FIB-induced damage layer. Uzuhashi and Ohkubo [18], for example, distinguish the fully amorphized layer from the total damaged region, the latter including material in which the lattice is distorted but still crystalline. In the present analysis, the amorphous–crystalline interface is assigned by the loss of the phase signal associated with laterally aligned Si atomic columns. This criterion is expected to approximate the boundary of the amorphized layer, although it may also include a small contribution from the adjacent distorted crystalline region. Our estimated thicknesses therefore generally lie close to, or slightly above, the previously reported amorphized-layer thicknesses.

The region milled at a final voltage of 30 kV shows both the largest deviation and the widest distribution of reported values. This can be attributed to the fitted interface positions varying substantially from column to column even within the narrow analyzed region (**Fig. S8d**), and to the structurally distorted region adjacent to the amorphous layer extending farther than at lower milling voltages. The greater total specimen thickness of this region (~68 nm) also presents a more demanding regime for conventional multislice reconstruction [40], and may contribute additional uncertainty to the recovered depth profile. Despite these limitations, the MEP reconstructions resolved and quantified the amorphous layers for all four milling conditions directly from the lamella used for atomic-resolution analysis.

## 4. Discussion

Interpreting the optimized thickness $t^*$ as an estimate of the physical thickness requires examining the relation between the multislice model and the specimen. In the multislice forward model, each slice is represented as an infinitesimally thin transmission function, separated from adjacent slices by a free-space propagation distance $\Delta z$, giving a total model thickness of $t = N \times \Delta z$. This model thickness is therefore a reconstruction parameter and need not correspond to the physical thickness of the specimen. Accordingly, a convergence alone does not establish such a correspondence. For example, atomically thin two-dimensional materials can be reconstructed successfully using multislice ptychography with slice stacks several times thicker than the specimen, where the atomic layers occupy only part of the slice stack and the remaining slices represent vacuum [41]. Reconstructing at a model thickness $t$ therefore says little about the physical thickness. The relevant question is under what conditions the error-minimizing thickness $t^*$ provides a reliable estimate of the physical specimen thickness.

Two features of the optimization distinguish the estimated thickness $t^*$ from a purely numerical reconstruction parameter. First, the optimization yields a well-localized joint minimum in thickness and defocus, consistent with the reported sensitivity of multislice reconstruction to its starting parameters. Second, the number of slices $N$ has a distinct physical and numerical role [42]. The slice spacing $\Delta z = t/N$ must be sufficiently small to capture dynamical scattering through the specimen, whereas excessively fine slicing makes the propagation between adjacent slices negligible, leaving the data unable to constrain the slices independently [42–44]. The optimum spacing depends not only on the diffraction data but also on the layer regularization. Because the regularization operates in the Fourier coordinates of the slice stack, its cutoff is defined per slice rather than as an absolute length, and the physical depth over which it smooths grows with both the regularization strength and the slice spacing. Maintaining a comparable smoothing depth under a stronger setting therefore requires finer slicing. The optimum $N^*$ is therefore associated with the effective depth sampling and regularization of the reconstruction, while the total thickness $t^*$ is substantially more robust to these settings.

The retrieved thickness remains subject to systematic uncertainties arising from both the reconstruction model and its experimental validation. For the three positions examined along the specimen prepared with a final milling voltage of 2 kV, the optimized thickness was 10–17% smaller than that obtained by the EELS log-ratio method. This offset is comparable to the 10–15% systematic underestimation reported for MEP depth sectioning where an elastic forward model is applied to diffraction patterns containing thermal diffuse scattering [45], suggesting one possible contribution to the discrepancy. The EELS reference itself also carries uncertainty because its absolute thickness depends on the assumed inelastic mean free path [37]. Layer regularization introduces an additional limitation. Filtering in the Fourier coordinates of the slice stack is periodic along the depth direction and can produce wrap-around artifacts that couple the opposite ends of the reconstructed volume, thereby blurring the surfaces that define $t^*$. Locating specimen terminations is a known challenge in multislice reconstructions [46], and real space depth regularization has been reported to suppress this artifact [47]. Further refinement of the forward model and depth regularization may therefore improve the accuracy with which specimen boundaries, and consequently specimen thickness, can be determined.

A practical limitation is that the optimization still requires physically reasonable search bounds. If the bounds are excessively broad, parameter combinations that reproduce the measured diffraction amplitudes without corresponding to the physical specimen geometry can become competitive. The present approach should therefore be regarded as a refinement of a coarse thickness estimate rather than a prior-free estimation. Such a prior estimate can be obtained using established methods, including electron-transparency contrast during FIB preparation [48], position-averaged CBED matching [9], or low-loss EELS [37]. In the present workflow, however, these methods are used only to define a plausible search range.

The local thickness used for the final multislice reconstruction is subsequently inferred from the same 4D-STEM dataset used for structural analysis.

## 5. Conclusions

We have demonstrated that the local thickness of a TEM lamella can be estimated from 4D-STEM data by coupling multislice electron ptychography with Bayesian optimization of the specimen thickness, probe defocus, and number of slices. Joint optimization is necessary because thickness and defocus partially compensate for one another, while their optimized values also depend on the number of slices. For the FIB-prepared silicon lamella examined here, the diffraction data produce a well-defined optimum in thickness and defocus, while varying the layer regularization shifts the preferred number of slices but leaves the optimized thickness comparatively stable. Moreover, independent optimization of different positions along a wedge-shaped region reproduces the expected increase in local thickness following the same spatial trend measured independently by EELS. Together, these observations support interpreting the optimized $t^*$ as an estimate of the local physical thickness rather than solely as a numerical reconstruction parameter, within the systematic uncertainties of the present model.

Application of these optimized parameters separates the FIB-induced amorphous surface layers from the crystalline silicon interiors with amorphous-layer thicknesses increasing monotonically with milling voltage and agreeing reasonably with previously reported conventional cross-sectional measurements. Unlike those measurements, however, our approach determines the damage directly from the same lamella and the same reconstructed slice stack used for atomic-resolution structural analysis, without requiring a separately prepared cross-sectional reference specimen.

More broadly, this procedure is particularly useful for quantitative 4D-STEM and multislice analyses in which dynamical scattering makes simulated and experimental signals strongly dependent on local thickness. Because it requires only the acquired diffraction data and a coarse prior estimate of the thickness, it can be incorporated into routine MEP workflows, offering a route to both specimen-specific reconstruction parameters and a quantitative assessment of the specimen quality itself.

## CRediT authorship contribution statement

K.-Y.P., under the guidance of C.S.C., designed the experiments, prepared the FIB lamella, performed multislice electron ptychography reconstructions, analyzed the data, and wrote the manuscript. Under the supervision of Y.-T.S., C.S. and M.T.P. acquired the 4D-STEM data. S.C. and S.L. contributed to the implementation of the Bayesian optimization pipeline and the thickness measurement code. J.O. performed the EELS measurements. C.S.C. supervised the project and acquired funding.

## Declaration of Competing Interest

The authors declare that they have no known competing financial interests or personal relationships that could have appeared to influence the work reported in this paper.

## Data availability

The 4D-STEM datasets, the reconstructed volumes, and the code used for the Bayesian optimization and the thickness analysis are available at https://doi.org/10.5281/zenodo.22106790.

## Acknowledgments

This work was supported by the Creative-Pioneering Researchers Program through Seoul National University (No. 3348-20260014), the National Research Foundation of Korea (NRF) grant funded by the Korean government (MSIT) (Grant No. RS-2024-00355333), and Samsung Electronics Co., Ltd. (IO250611-13054-01). The microscopy work at USC was supported by the U.S. Department of Energy, Basic Energy Sciences (award No. DE-SC0025423). Data processing was carried out using the cSAXS ptychography MATLAB package developed by the Science IT and the coherent X-ray scattering (CXS) groups, Paul Scherrer Institut, Switzerland. The tilt-propagator extension to fold_slice was developed by L. Bhatt (Muller Group, Cornell University) [30].

## Declaration of generative AI and AI-assisted technologies

The authors used ChatGPT (OpenAI) to assist with English translation and improve the language and readability of the manuscript. All AI-assisted text was subsequently reviewed and revised by the authors, who accept full responsibility for the final content.

# Figures

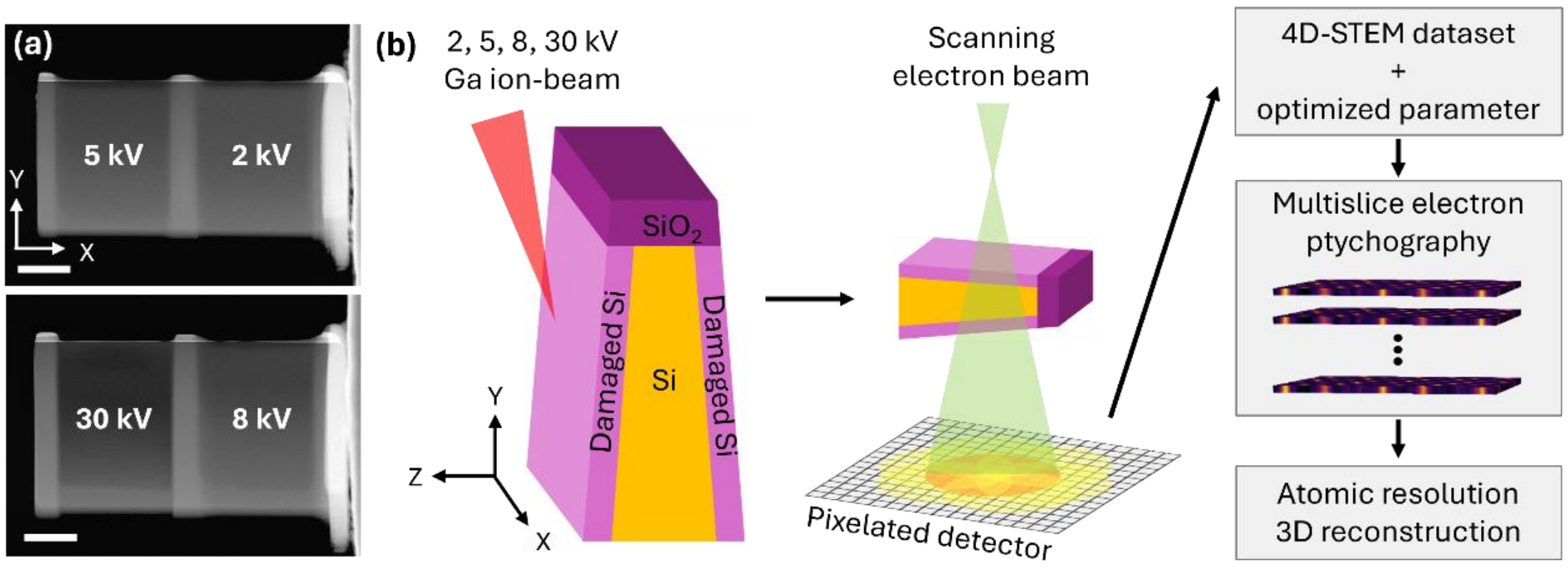

**Figure 1. Specimen geometry and experimental workflow for multislice electron ptychography.** (a) STEM images showing four regions in the silicon lamella, milled with final milling voltages of 2, 5, 8, and 30 kV. Scale bars, 2 µm. (b) Schematic of the FIB-prepared cross-sectional lamella and the 4D-STEM acquisition along with reconstruction workflow. The specimen is milled from both sides with the Ga ion beam, which leaves FIB-induced amorphous layers on both surfaces. A focused electron probe is then scanned across the lamella while a diffraction pattern is recorded at each probe position, forming a 4D-STEM dataset. The specimen thickness and reconstruction parameters are optimized from the diffraction data and subsequently used for multislice electron ptychography, allowing estimation of the total specimen thickness as well as atomic-resolution, depth-resolved reconstruction of the lamella.

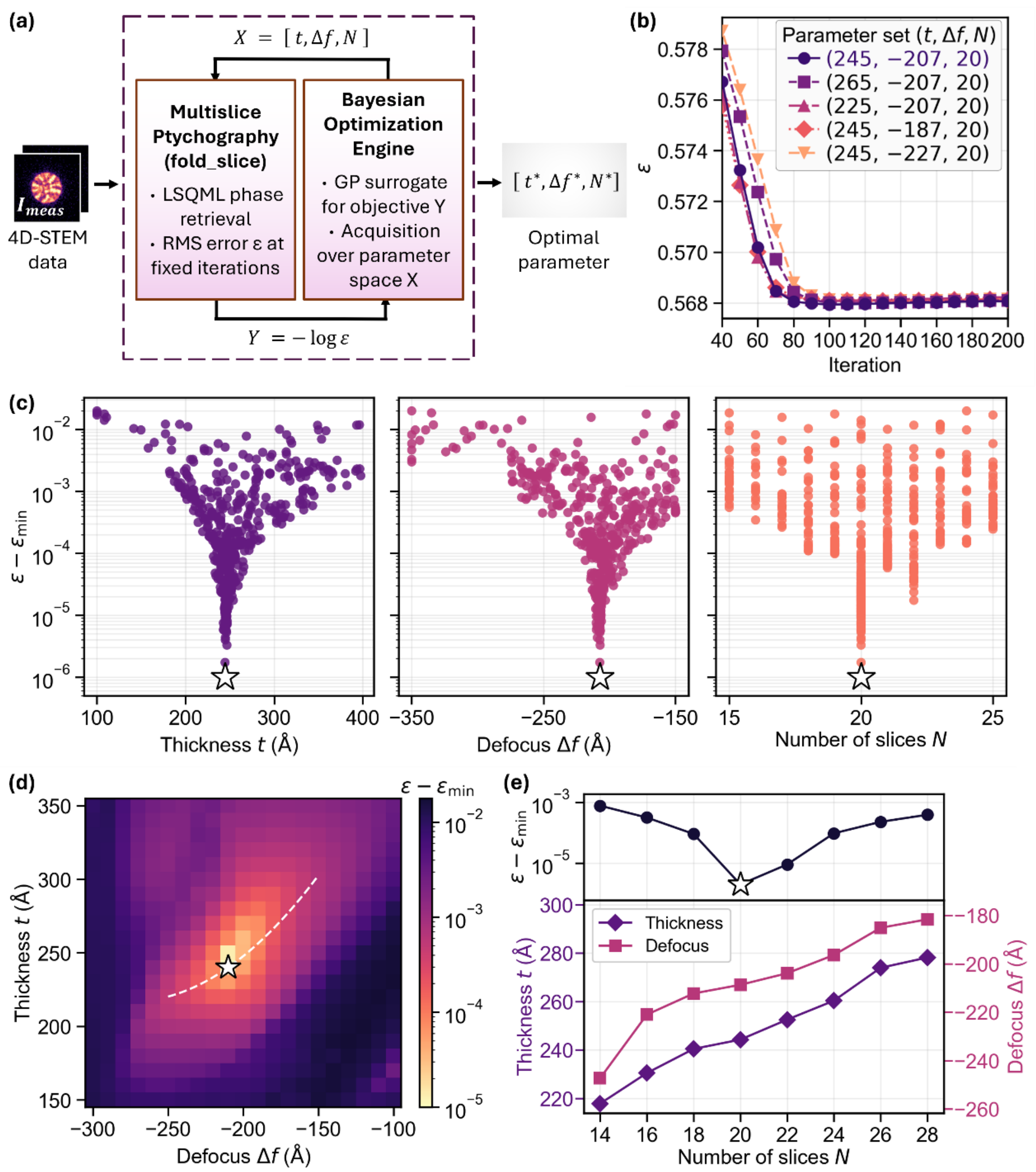

**Figure 2. Bayesian optimization for joint estimation of specimen thickness and reconstruction parameters.** All panels show results from the 2 kV region. (a) Optimization workflow. Each candidate parameter set $X = [t, \Delta f, N]$ is evaluated by a multislice ptychographic reconstruction, yielding the objective $Y = -\log \varepsilon$, which is used to fit a Gaussian process (GP) surrogate. Here, $t$ is the specimen thickness, $\Delta f$ the probe defocus, $N$ the number of slices, and $\varepsilon$ the RMS error between measured and simulated amplitudes. $I_{meas}$ is the measured diffraction intensity and LSQML stands for least-squares maximum-likelihood. (b) RMS error as a function of reconstruction iteration for the optimum and four sets perturbed by ±20 Å in $t$ and $\Delta f$. All reach a plateau beyond approximately one hundred iterations, and their relative ranking is preserved. (c) Excess error, $\varepsilon - \varepsilon_{min}$, of all candidates from the joint optimization, each plotted against $t$, $\Delta f$, and $N$. (d) Excess error landscape in the $t$–$\Delta f$ plane, computed on a grid at fixed $N^* = 20$. The dashed line traces the error-minimizing

thickness at each defocus, illustrating the coupling between $t$ and $\Delta f$. (e) Minimum excess error obtained with $N$ fixed at each value (top) and the corresponding optimized $t$ and $\Delta f$ (bottom), demonstrating their systematic dependence on $N$. Stars in (c–e) indicate the joint optimum $X^* = [245\ \text{Å}, -207\ \text{Å}, 20]$.

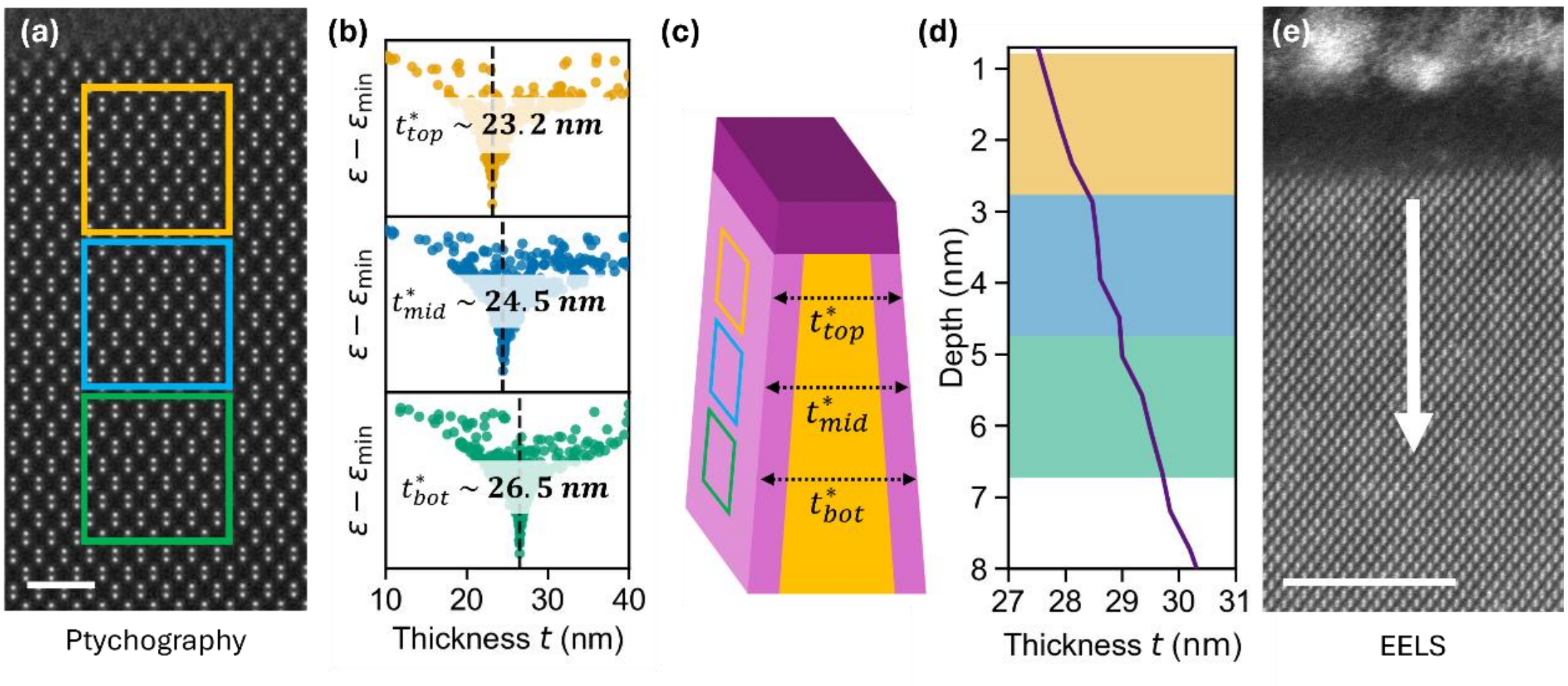

**Figure 3. Local thickness variation estimates from optimization and comparison with independent measurement by EELS.** All panels correspond to the 2 kV milled region. (a) Reconstructed phase image showing the top (yellow), middle (blue), and bottom (green) subregions. Scale bar, 1 nm. (b) Excess error of evaluated candidates as a function of thickness for each subregion as in Fig. 2c, yielding optimized thickness $t^*$. (c) Schematic of the wedge-shaped lamella showing the three subregions marked in (a). (d) Local specimen thickness measured independently by the EELS log-ratio method along the direction indicated in (e), with the depth ranges corresponding to the three subregions shaded in the matching colors as in (a–c). (e) Annular dark-field (ADF) STEM image acquired during EELS measurement, with the arrow indicating the direction of the thickness profile in (d). Scale bar, 5 nm.

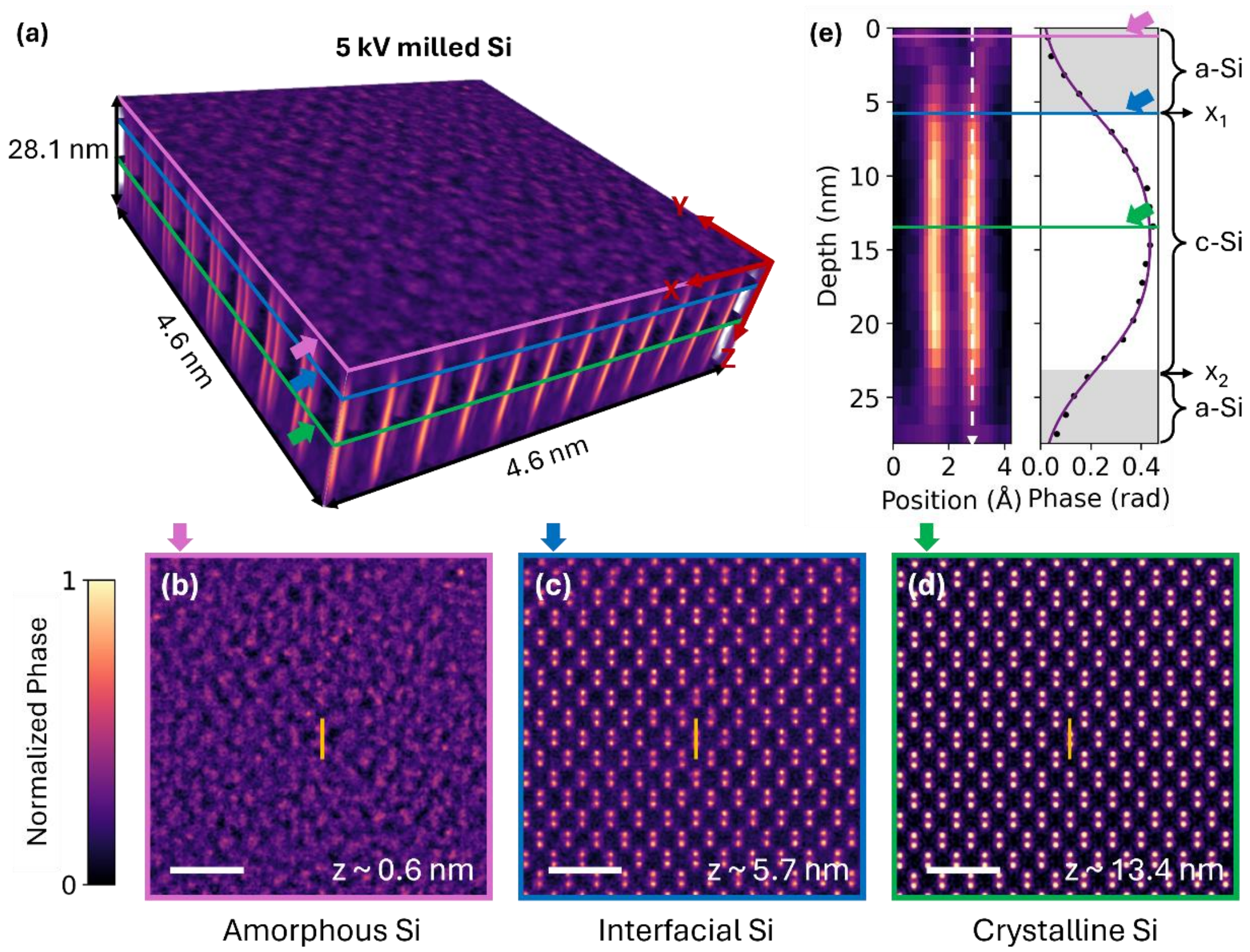

**Figure 4. Depth-resolved separation of amorphous and crystalline regions in the 5 kV milled silicon lamella.** (a) Three-dimensional rendering of the reconstructed phase volume, with the depths of the slices in (b–d) marked in the matching colors. (b–d) Lateral slices extracted at depths of z ~ 0.6, 5.7, and 13.4 nm, each corresponding to (b) the disordered amorphous region, (c) the amorphous–crystalline interface, and (d) the fully resolved crystalline bulk, respectively. Scale bars, 1 nm. (e) Phase-versus-depth profile of the silicon dumbbell indicated by the orange line in (b–d). The left panel shows the phase as a function of lateral position and depth, revealing the persistence of the two atomic columns through the crystalline interior and their disappearance toward the specimen surfaces. The right panel shows the depth profile extracted at a single column marked by the white dashed line, together with the fitted double error function. The fitted half-height positions $x_1$ and $x_2$ define the amorphous–crystalline interfaces separating the crystalline bulk (c-Si) from the amorphous silicon layers (a-Si), shaded in grey, on either side.

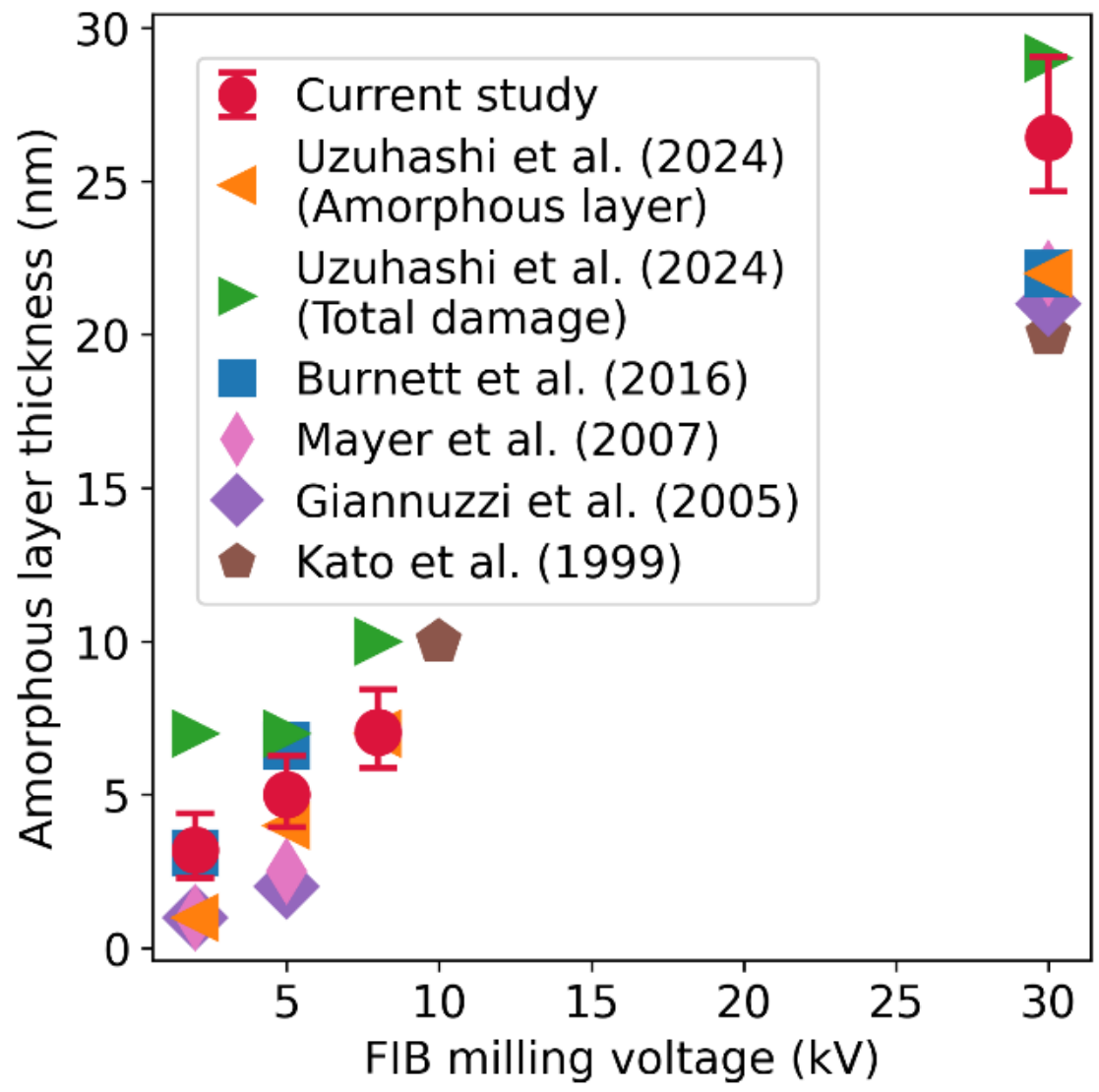

**Figure 5. FIB-induced amorphous-layer thickness as a function of final milling voltage.** Amorphous-layer thicknesses determined from the depth-resolved MEP reconstructions are shown for final FIB milling voltages of 2, 5, 8, and 30 kV. The MEP-based estimates (red) in this study correspond to the mean thickness of a single amorphous layer, with error bars spanning the values obtained from individual column profiles. Literature values obtained from a single surface in cross-sectional measurements are shown for comparison [11–13,16,18]; Uzuhashi and Ohkubo report both the fully amorphized layer and the total damaged region, which are plotted separately.

# Supplementary Materials for:

# Optimization-Based Thickness Estimation and Depth-Resolved FIB-Induced Damage Characterization via Multislice Electron Ptychography

Keun-Yeol Park[1], Chuqiao Shi[2], Murat Tuna Pamuk[2], Sooyoung Cheong[1], Seonyu Lee[1], Juhui Oh[1], Yu-Tsun Shao[2,3], and Celesta S. Chang[1,4,*]

[1] *Department of Physics and Astronomy, Seoul National University, Seoul 08826, Korea*

[2] *Mork Family Department of Chemical Engineering and Materials Science, University of Southern California, Los Angeles, CA 90089, USA*

[3] *Core Center of Excellence in Nano Imaging, University of Southern California, Los Angeles, CA 90089, USA*

[4] *Institute of Applied Physics, Seoul National University, Seoul 08826, Korea*

* *Corresponding author: celesta@snu.ac.kr*

**Supplementary Table 1. FIB milling conditions for the four regions finalized at 2, 5, 8, and 30 kV.**

| Specimen region | **2 kV** | **5 kV** | **8 kV** | **30 kV** |
|---|---|---|---|---|
| Milling parameters[a] | Thickness down to[b] | | | |
| 30 kV, 0.44 nA, CCS, ±1.2° | 1 μm | | | |
| 30 kV, 0.26 nA, CCS, ±1.2° | 500 nm | | | |
| 30 kV, 90 pA, CCS, ±1.2° | 200 nm | | | |
| 30 kV, 90 pA, rectangle, ±3° | 100 nm | 100 nm | 100 nm | ~60 nm† |
| 8 kV, 61 pA, rectangle, ±3° | - | - | ~30 nm* | - |
| 5 kV, 63 pA, rectangle, ±3° | ~60 nm† | ~30 nm* | - | - |
| 2 kV, 66 pA, rectangle, ±3° | ~30 nm* | - | - | - |

CCS: Cleaning cross section

a: Ion beam accelerating voltage, beam current, scanning pattern, and tilt relative to the lamella

b: Unmarked thickness values are rough estimates from direct imaging in the dual-beam FIB-SEM

†,*: Estimated from the relative brightness of the lamella in secondary-electron SEM images acquired with an Everhart–Thornley detector, at 3 kV for the ~60 nm values (†) and 2 kV for the ~30 nm values (*). This is a common qualitative endpoint criterion in FIB thinning (A.P. Conlan, E. Tillotson, A. Rakowski, D. Cooper, S.J. Haigh, Direct measurement of TEM lamella thickness in FIB-SEM, J. Microsc. 279 (2020) 168–176, https://doi.org/10.1111/jmi.12852).

**Supplementary Table 2. Experimental acquisition conditions and optimized reconstruction parameters for the four FIB-milled regions.**

| Specimen region | **2 kV** | **5 kV** | **8 kV** | **30 kV** |
|---|---|---|---|---|
| Accelerating voltage (kV) | 200 | | | |
| Probe semi-angle (mrad) | 30 | | | |
| Scan dwell time (μs) | 35 | | | |
| Electron dose ($e^-/Å^2$) | $7 \times 10^4$ | | | |
| Scan pixel (px) | 256 × 256 | | | |
| Scan step size (Å) | 0.36 | 0.36 | 0.36 | 0.35 |
| Detector pixel (px) | 192 × 192 | | | |
| Diffraction sampling (mrad) | 0.81 | 0.81 | 0.81 | 0.79 |
| Collection angle (mrad) | 78 | 78 | 78 | 76 |
| Layer regularization | 0.2 | | | |
| Model diffraction blur σ (px) | 1 | | | |
| Number of probe modes | 1 | | | |
| Batch size | 64 | | | |
| Number of iterations[a] | 1000 | | | |
| Scan rotation (°) | 0.3 | 1.0 | 1.3 | 0.1 |
| (x, y) tilts (mrad) | (2, 0) | (3, -1) | (4, 0) | (5, 3) |
| Thickness (Å) | 245 | 281 | 336 | 680 |
| Defocus (Å)[b] | −207 | −170 | −200 | 175 |
| Number of slices | 20 | 22 | 23 | 31 |
| Slice spacing (Å) | 12.3 | 12.8 | 14.6 | 21.9 |

a: For the final reconstruction after parameter optimization.

b: Positive and negative defocus values correspond to underfocus and overfocus, respectively.

**Supplementary Table 3. Bayesian optimization settings and search bounds.**

| Specimen region | **2 kV** | **5 kV** | **8 kV** | **30 kV** |
|---|---|---|---|---|
| Surrogate model | Gaussian process (RBF kernel with ARD) | | | |
| Input transform | Normalized to $[0, 1]^3$ | | | |
| Integer handling | Rounding transform on N | | | |
| Acquisition function | qUCB ($\beta = 0.1$), 512 QMC samples | | | |
| Batch size q | 4 | | | |
| Acquisition optimizer | 20 restarts, 1024 raw samples | | | |
| Initial random evaluations | 20 | | | |
| Total evaluations | ≥ 500 | | | |
| Iterations per candidate | 100 | | | |
| Thickness bounds (Å) | [100, 400] | [100, 400] | [200, 500] | [500, 800] |
| Defocus bounds (Å) | [−350, −150] | [−300, −100] | [−300, −100] | [100, 300] |
| Number of slices bounds | [15, 25] | [15, 25] | [20, 30] | [25, 35] |

RBF: radial basis function

ARD: automatic relevance determination

qUCB: q-batch upper confidence bound

QMC: Quasi-Monte Carlo

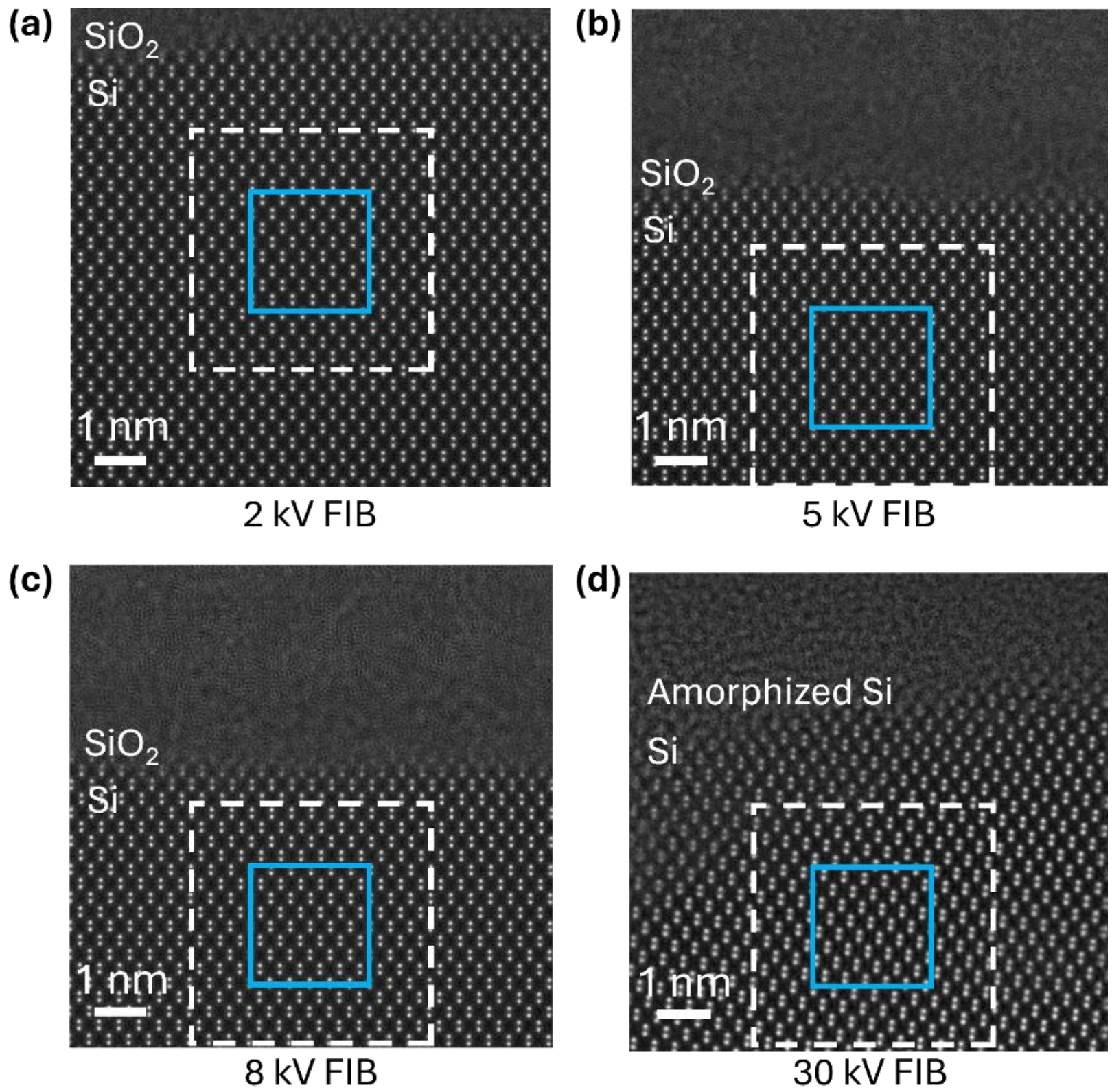

**Supplementary Figure 1. Reconstructed central slices of the four milled regions and the regions selected for optimization and depth-resolved analysis.** (a–d) Central slices of the reconstructed volumes of the 2, 5, 8, and 30 kV milled regions, respectively, each spanning the full $9.2 \times 9.2$ nm² scan field of view. Blue solid boxes indicate the $64 \times 64$ scan pixel ROIs ($2.3 \times 2.3$ nm²) cropped for Bayesian optimization of the specimen thickness, defocus, and the number of slices. The optimized parameters were then used to reconstruct the full fields of view shown in each panel and to quantify amorphous-layer thicknesses at individual atomic columns within the ROIs. White dashed boxes indicate the $4.6 \times 4.6$ nm² regions, centered on the optimization ROIs, used for the three-dimensional renderings and depth-resolved analysis in Fig. 4a, S8 and S9. In (a), the blue ROI corresponds to the middle subregion in Fig. 3a. In (d), the upper field of view shows amorphous silicon rather than the original $SiO_2$ surface layer because FIB-induced amorphous layers generated at the two milled surfaces meet at the thinnest part of the wedge, leaving no crystalline silicon between them (Fig. S2).

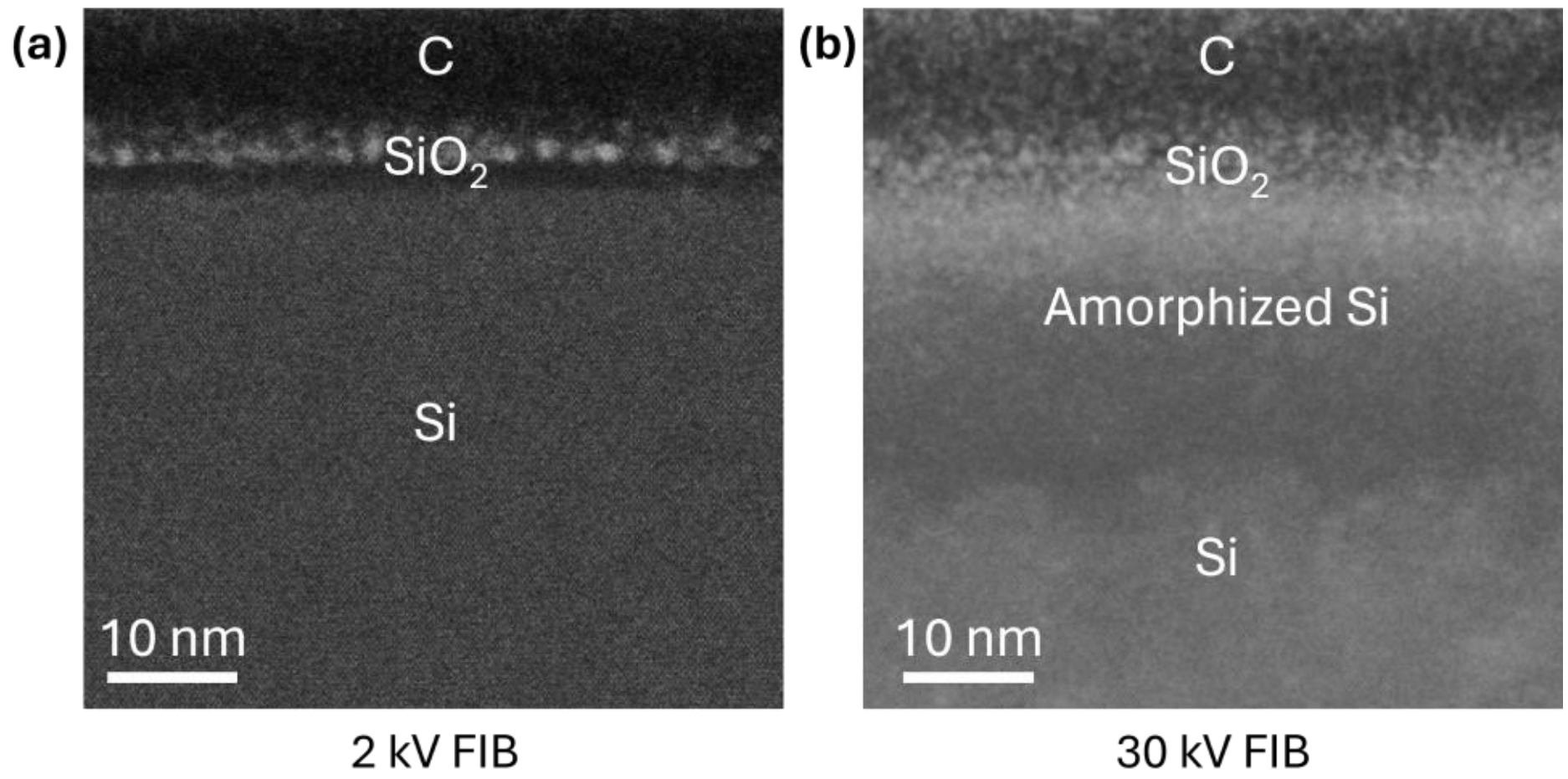

**Supplementary Figure 2. HAADF-STEM images showing the distinct near-surface structures of the 2 and 30 kV FIB-milled regions.** High-angle annular dark-field scanning transmission electron microscopy (HAADF-STEM) images of the lamella regions finalized at (a) 2 and (b) 30 kV. In the 2 kV region, the carbon (C) protection layer and the original $SiO_2$ layer sit directly above crystalline silicon. The bright region between the C and $SiO_2$ layers consists of clusters formed by the reaction of implanted Ga with oxygen during milling. In the 30 kV region, amorphized silicon separates the $SiO_2$ from the crystalline core. Toward the thinner, upper part of the 30 kV wedge, amorphous layers generated from the two milled surfaces extend through the full local specimen thickness, leaving no crystalline silicon in projection. Crystalline silicon therefore appears only in the thicker, lower part of the lamella. The reconstructed central slices of the 5 and 8 kV regions likewise show the $SiO_2$ directly adjacent to crystalline silicon as in the 2 kV region (Fig. S1b, c).

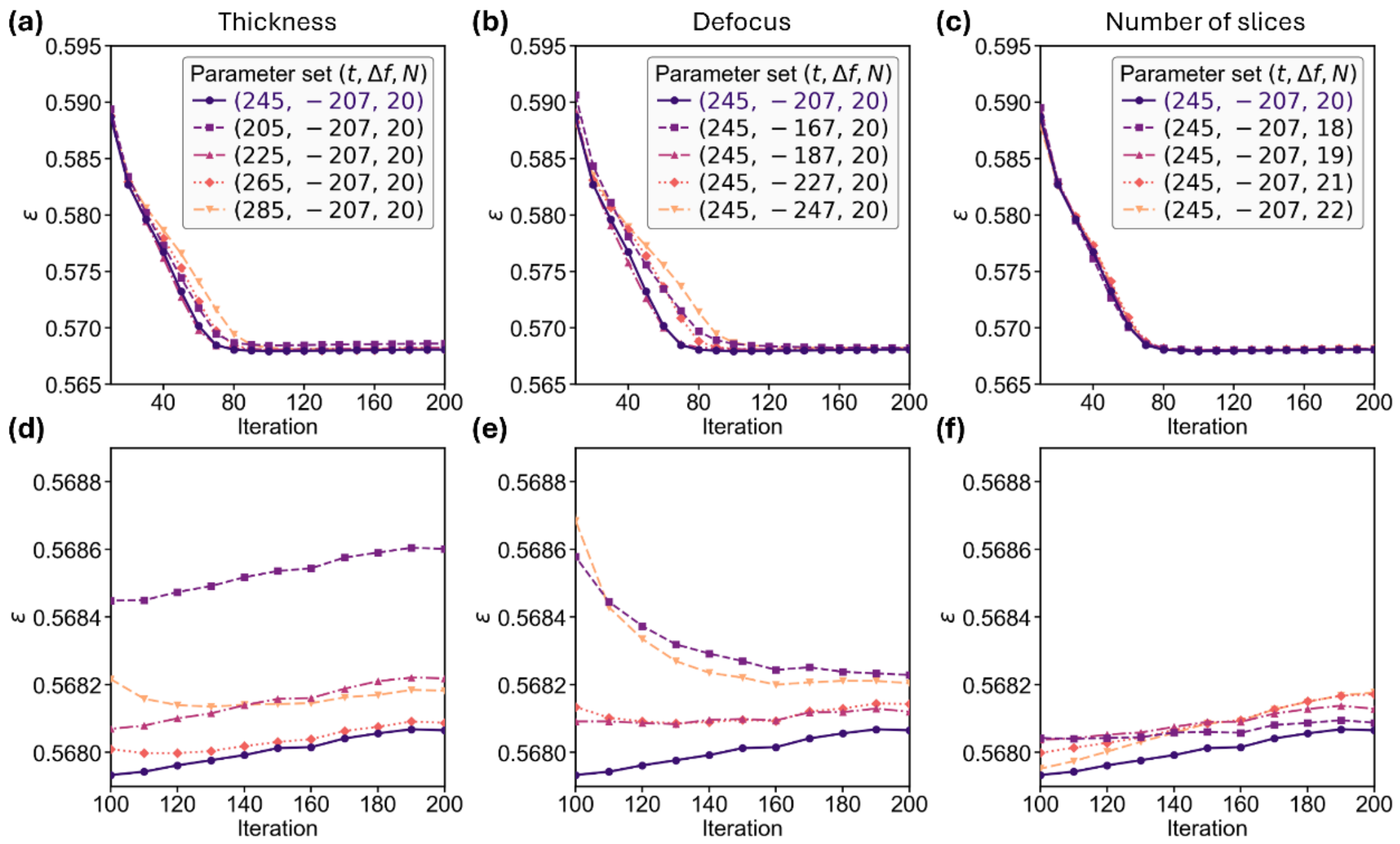

**Supplementary Figure 3. Reconstruction-error convergence and ranking stability under controlled parameter perturbations.** RMS error ε as a function of iteration number for parameter sets obtained by perturbing one parameter at a time about the optimum for the 2 kV region, $X^* = [t^*, \Delta f^*, N^*] = [245\ \text{Å}, -207\ \text{Å}, 20]$. (a–c) RMS error trajectories over 200 iterations for perturbations in (a) specimen thickness (±20 and ±40 Å), (b) probe defocus (±20 and ±40 Å), and (c) number of slices (±1 and ±2) while the remaining two parameters were held at their optimal values. All parameter sets reach a plateau beyond approximately 100 iterations. (d–f) Enlarged views of the plateau regions in (a–c). Although $\varepsilon$ varies slightly beyond 100 iterations, the optimized parameter set remains the lowest-error solution, because once the data-fidelity improvement has saturated, the residual change in ε is dominated by the normalization of the probe and object applied during the reconstruction rather than by further refinement of the fit. The relative ranking of the perturbed sets is preserved, justifying the fixed 100-iteration budget used to evaluate candidates during Bayesian optimization.

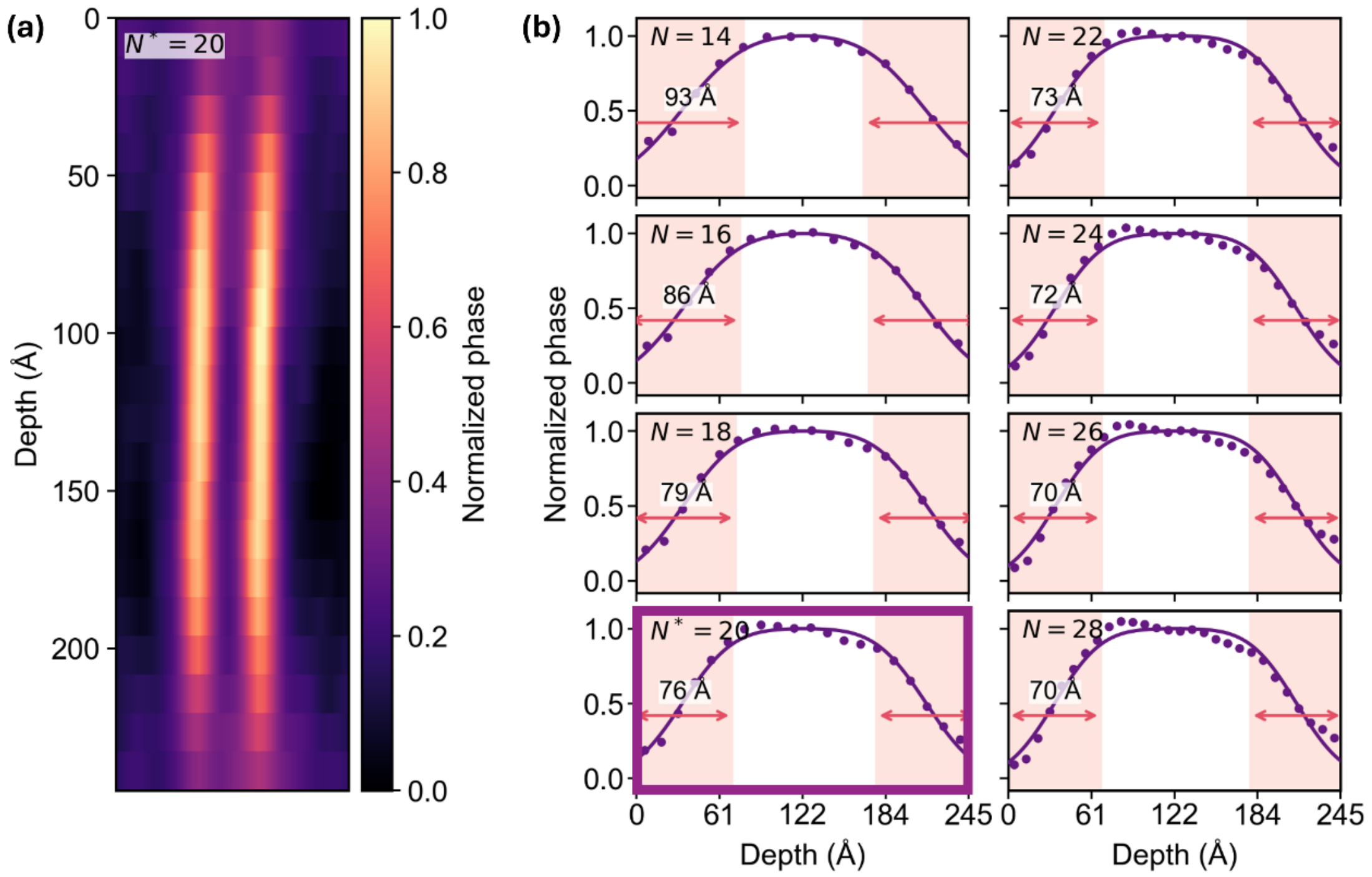

**Supplementary Figure 4. Dependence of depth resolution on the number of slices.** (a) Depth section through the reconstructed 2 kV region at the optimized number of slices, $N^*$ = 20, showing the phase of two crystalline silicon atomic columns extending through the specimen. (b) Phase-versus-depth profiles extracted along the right-hand column in (a) for reconstructions with $N$ = 14–28. The total thickness and probe defocus were fixed at their optimized values, $t^*$ = 245 Å and $\Delta f^*$ = −207 Å, varying axial discretization $\Delta z = t^*/N$. Points are the reconstructed phase, and solid lines are symmetric double-error-function fits. Shaded regions with arrows indicate the transition width at each interface, taken as the 10–90% rise of the fitted profiles, with the corresponding values given in each panel. The transition width decreases rapidly up to $N$ = 18, changes only slightly at $N$ = 20 and 22, and remains nearly constant thereafter. Thus, finer slicing beyond $N \approx 20$ yields progressively less improvement in depth resolution. The panel outlined in color is the optimum identified independently by Bayesian optimization, $N^*$ = 20, at the onset of this plateau.

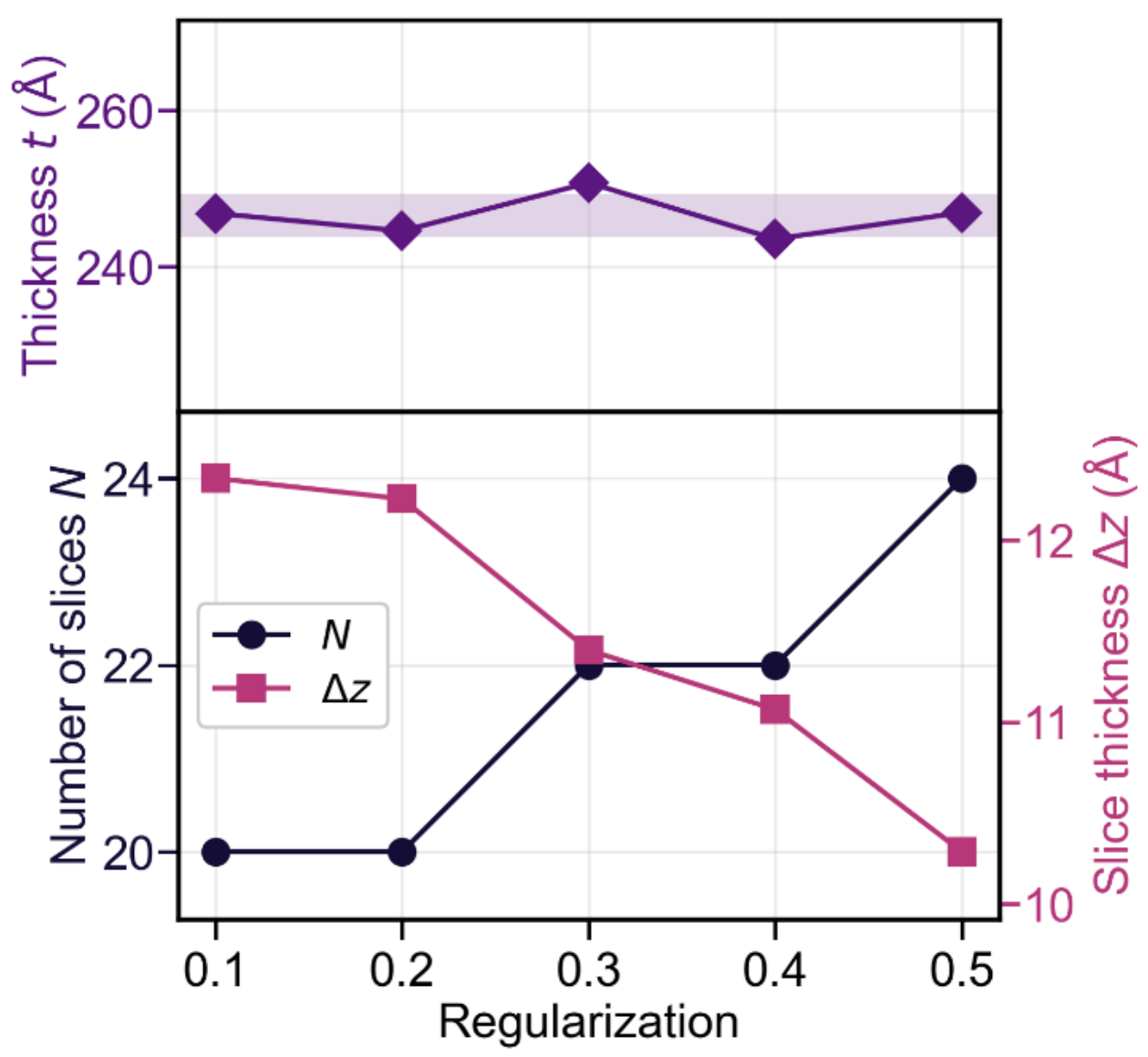

**Supplementary Figure 5. Dependence of the optimized parameters on the layer regularization.** Joint Bayesian optimization of the 2 kV region with the layer regularization varied from 0.1 to 0.5. All other reconstructions reported in this work used 0.2. The optimized thickness (top) stays within 3% of $t^*$ = 245 Å (shaded band) over the whole range. The optimized number of slices $N^*$ (bottom) increases from 20 to 24 as the regularization increases, and the slice spacing decreases correspondingly from 12.3 to 10.3 Å.

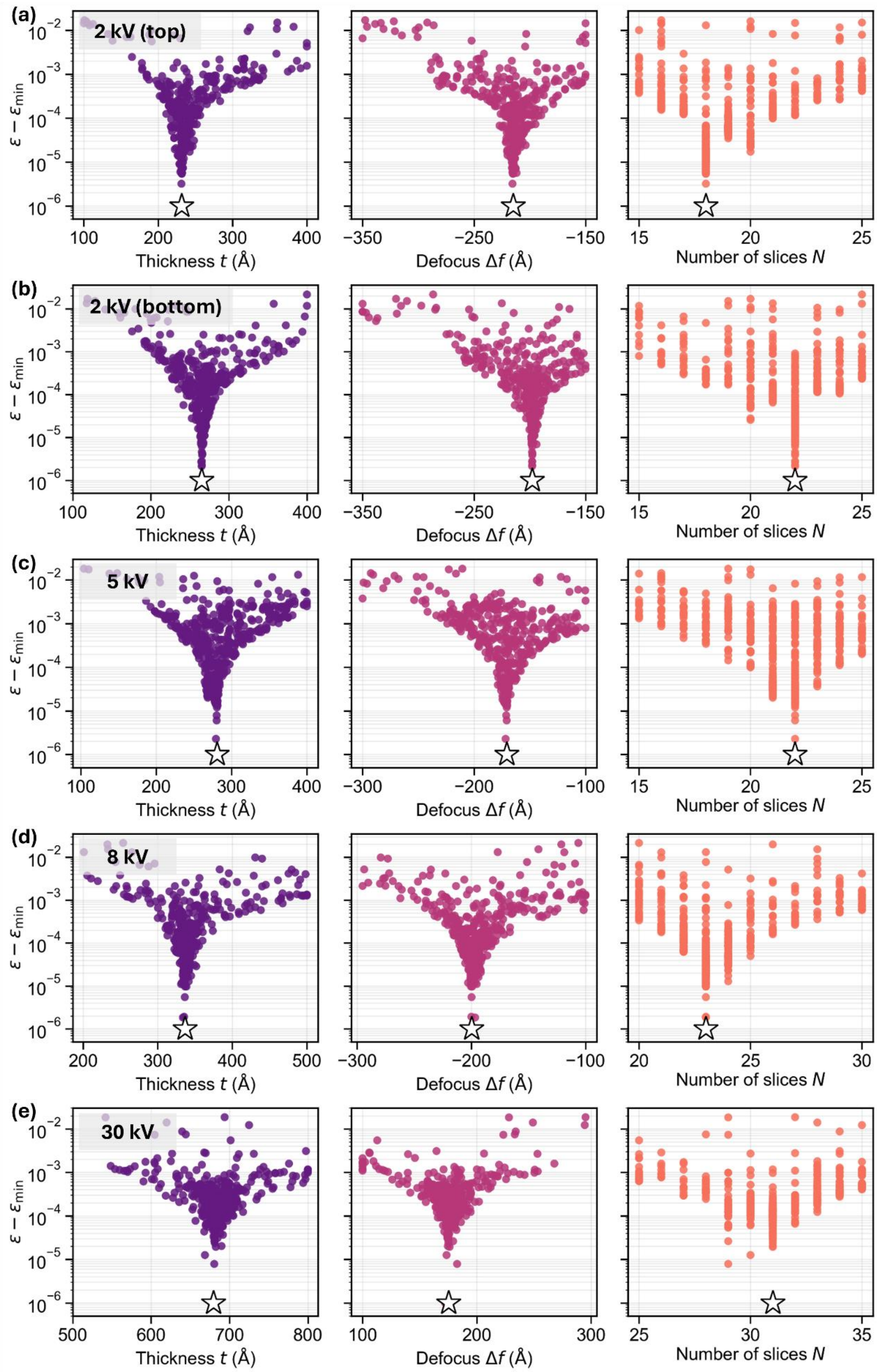

**Supplementary Figure 6. Joint parameter optimization for other subregions and milling voltages.** Excess error $\varepsilon - \varepsilon_{min}$ of all candidates from a single joint optimization, plotted against thickness, defocus, and number of slices as in Fig. 2c.

Stars indicate the optimized parameter in each case. (a, b) Top and bottom subregions of the 2 kV dataset, corresponding to the yellow and green ROIs in Fig. 3a, respectively, with the optimized thicknesses given in Fig. 3b. The corresponding data for the middle subregion are shown in Fig. 2c. (c–e) Optimization results for the 5, 8, and 30 kV regions, respectively, with the optimized values summarized in Table S2. The search bounds were chosen separately for each dataset to bracket the expected values, setting the axis limits of each panel. The corresponding ranges are listed in Table S3, and in all cases the optimum lies in the interior of the explored parameter ranges.

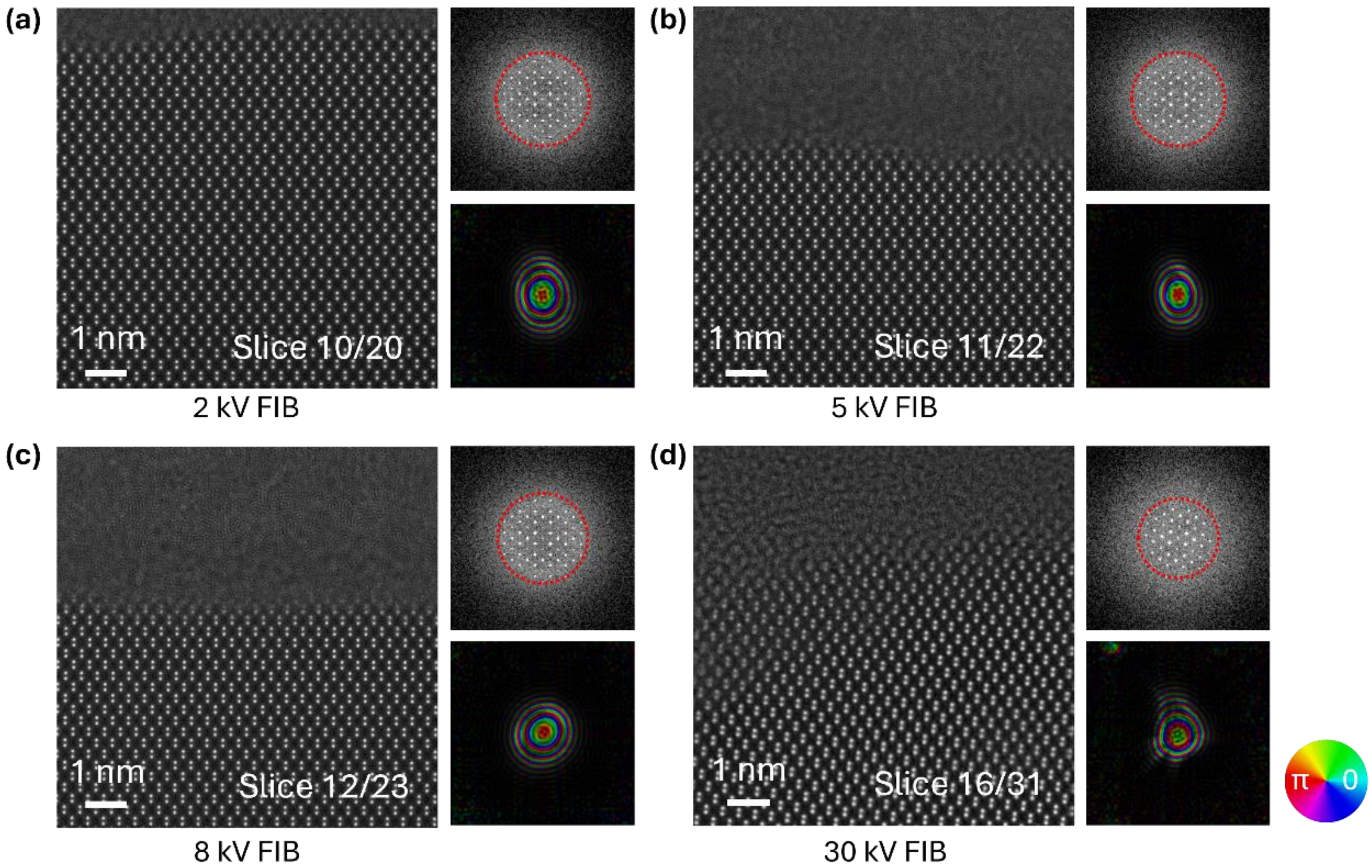

**Supplementary Figure 7. Reconstruction quality and spatial resolution of the four milled regions.** (a–d) Results from 2, 5, 8, and 30 kV regions, respectively. Each panel shows the phase image of the central reconstructed slice (left), its Fourier transform (top right), and the reconstructed probe at the specimen entrance surface (bottom right). The red dashed circles in the Fourier transforms pass through the highest-order Bragg reflections that remain distinguishable above the background, corresponding to spatial resolutions of 0.63, 0.63, 0.65, and 0.75 Å for (a–d), respectively. The reconstructed probes are displayed with amplitude represented by brightness and phase by hue, as indicated by the color wheel at lower right. All four reconstructions resolve the Si lattice at sub-Ångström resolution, with a modest reduction in spatial resolution for the thicker 30 kV region.

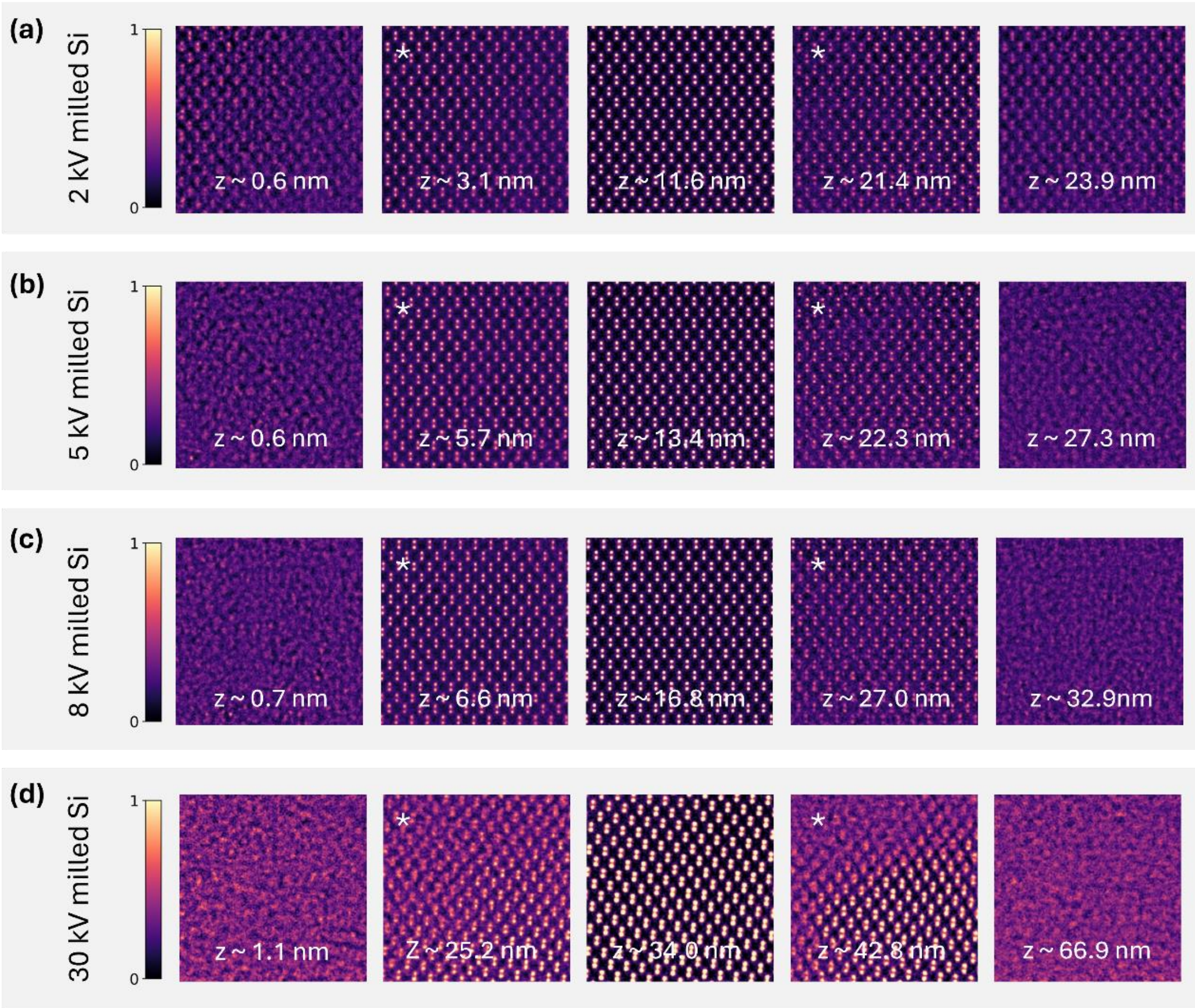

**Supplementary Figure 8. Depth-resolved evolution from amorphous surface layers to crystalline silicon.** (a–d) Reconstructed 2, 5, 8, and 30 kV regions, respectively, shown at five representative depths z labeled in each panel. Within each region all slices are displayed using the same phase scale normalized over the entire reconstructed volume, allowing direct comparison across depth. The outermost images were taken near the two ends of the reconstructed volume, while the central slice lies within the crystalline bulk. Asterisks indicate slices near the fitted amorphous–crystalline interface positions defined in Section 2.4, analogous to the interfacial slice as shown in Fig. 4c. At these slices, some atomic columns were already well defined while others remain indistinguishable from the disordered background, reflecting the column-to-column spread in the fitted interface positions. This variation was most pronounced in the 30 kV region (d). All four regions show the same progression from amorphous to crystalline and back, with the depth extent of the amorphous regions increasing with the final milling voltage.

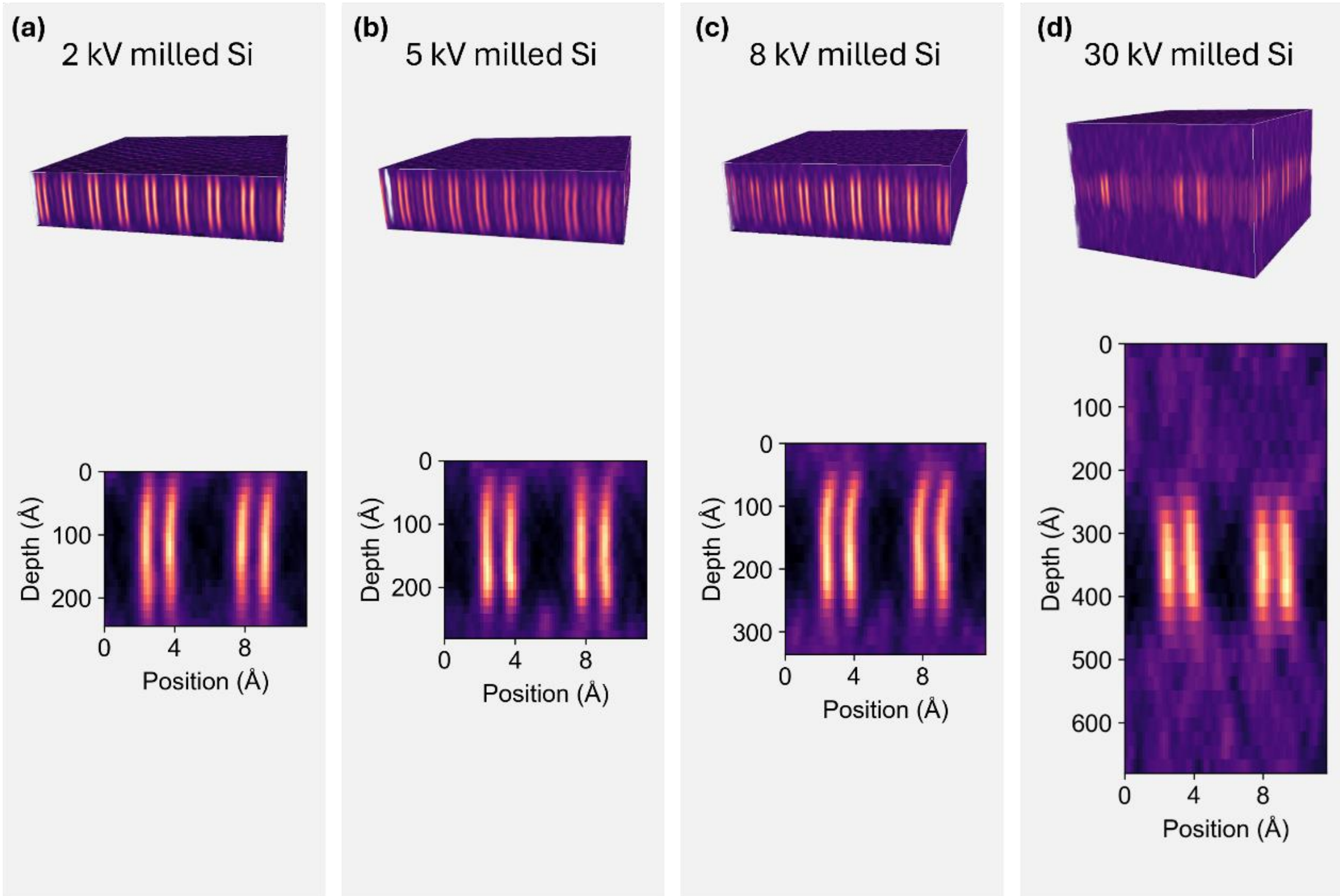

**Supplementary Figure 9. Three-dimensional phase renderings and atomic column depth profiles of regions finalized at 2, 5, 8, and 30 kV.** (a–d) Results from the 2, 5, 8, and 30 kV regions, respectively. For each region, the top panel shows the three-dimensional rendering of the reconstructed phase volume over the $4.6 \times 4.6\ nm^2$ region indicated in Fig. S1, analogous to Fig. 4a. The bottom panel shows the depth profile of a silicon dumbbell, as in Fig. 4e. The fraction of the reconstructed volume occupied by the crystalline interior decreases with increasing final milling voltage, consistent with the increasing amorphous-layer thickness quantified in Fig. 5.

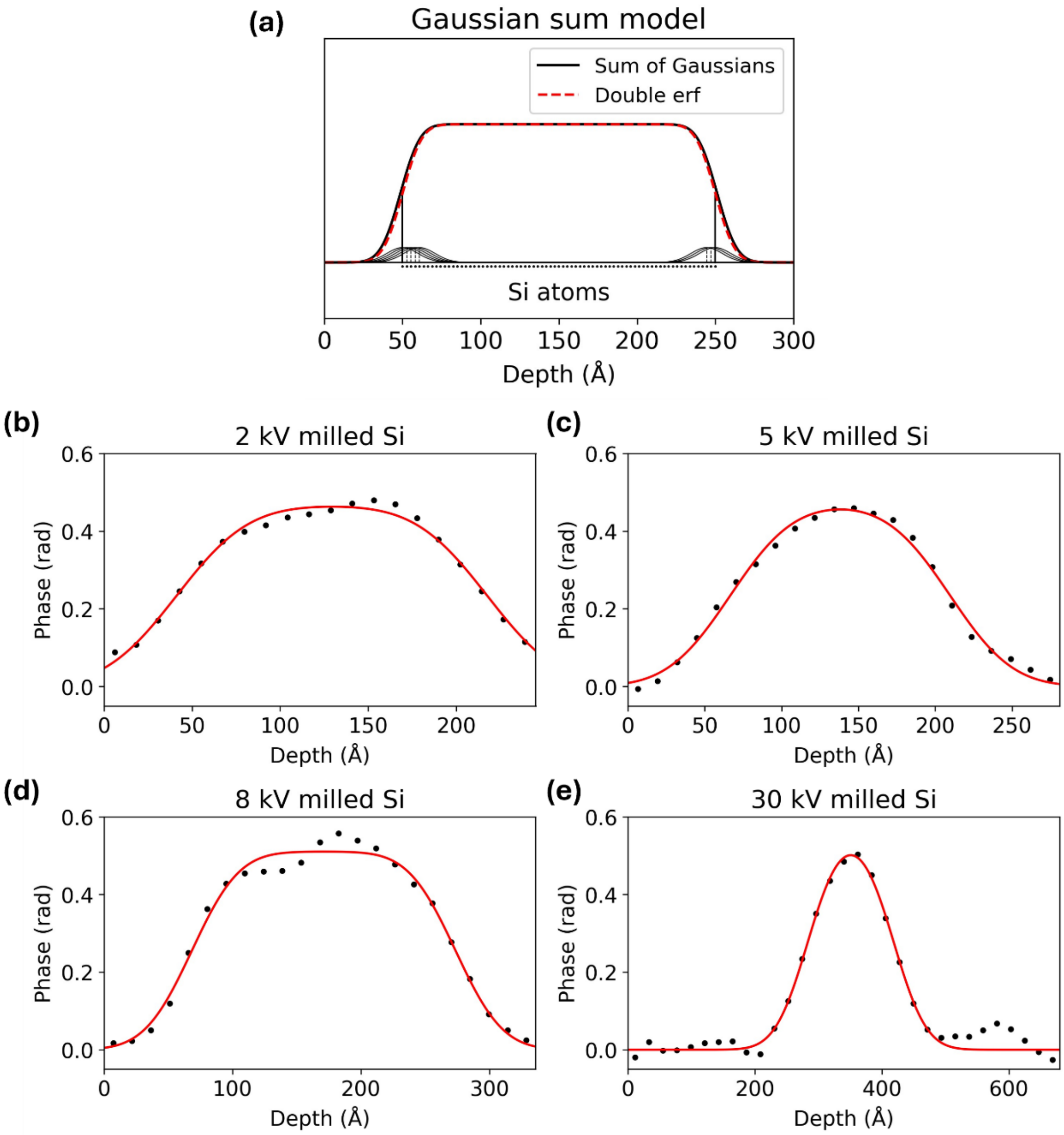

**Supplementary Figure 10. Double-error-function model for determining amorphous–crystalline interfaces from atomic column depth profiles.** (a) Schematic motivation for the fitting model used for profiling atomic columns. Individual silicon atoms are modeled with a Gaussian profile, and an atomic column is therefore a sum of closely spaced Gaussians. Their envelope is approximated by a symmetric double error function, providing a compact description of the two transitions between the crystalline interior and the disordered surface regions. The two interface positions are defined as the half-height positions of the corresponding error-function transitions. (b–e) Representative reconstructed phase versus depth profiles from the individual atomic columns in the 2, 5, 8, and 30 kV regions, respectively. Points show the reconstructed phase and red curves indicate the fitted symmetric double error functions.